\documentclass[11pt,a4paper]{article}
\pdfoutput=1
\usepackage{jheppub}

\usepackage{color}
\usepackage{feynmp}
\usepackage{multirow}
\usepackage{slashed}
\usepackage{mdframed}

\usepackage{natbib}
\usepackage{graphics}
\usepackage{amsmath,amsfonts}
\usepackage{epsfig}

\newcommand{\tr}{\operatorname{tr}}
\newcommand{\M}{\mathcal{M}}
\newcommand{\s}{\slashed}
\newcommand{\li}{\operatorname{Li}}

\renewcommand\u[1]{\text{#1}} % No space
\newcommand{\us}[1]{\text{ #1}} % Space

\newcommand{\xsecff}{(\sigma v)_{\chi\chi\to f\bar{f}}}
\newcommand{\xsecll}[1]{(\sigma v)_{\chi\chi\to {#1} {#1}}}
\newcommand{\xsecgg}{(\sigma v)_{\chi\chi\to \gamma\gamma}}
\newcommand{\xsecgx}[1]{(\sigma v)_{\chi\chi\to \gamma {#1}}}
\newcommand{\rxf}[2]{R_{{#1},{#2}}}%(M_\chi)}

\newcommand{\biL}[2]{[\bar{\chi}{#1}\chi][\bar{\tau}{#2}\tau]}

\begin{document}

\title{Monochromatic Gamma Rays from Dark Matter Annihilation to Leptons}

\preprint{SCIPP 15/06}

\author[a,b]{Adam Coogan}
\emailAdd{acoogan@ucsc.edu}

\author[a,b]{Stefano Profumo}
\emailAdd{profumo@ucsc.edu}

\author[a,b]{William Shepherd}
\emailAdd{wshepher@ucsc.edu}

\affiliation[a]{Department of Physics, University of California, 1156 High St., Santa Cruz, CA 95064, USA}
\affiliation[b]{Santa Cruz Institute for Particle Physics, Santa Cruz, CA 95064, USA} 

\abstract{We investigate the relation between the annihilation of dark matter (DM) particles into lepton pairs and into 2-body final states including one or two photons.  We parametrize the DM interactions with leptons in terms of contact interactions, and calculate the loop-level annihilation into monochromatic gamma rays, specifically computing the ratio of the DM annihilation cross sections into two gamma rays versus lepton pairs.  While the loop-level processes are generically suppressed in comparison with the tree-level annihilation into leptons, we find that some choices for the mediator spin and coupling structure lead to large branching fractions into gamma-ray lines. This result has implications for a dark matter contribution to the AMS-02 positron excess.  We also explore the possibility of mediators which are charged under a dark symmetry and find that, for these loop-level processes, an effective field theory description is accurate for DM masses up to about half the mediator mass.}


\arxivnumber{1504.05187}

\maketitle

\section{Introduction}
\label{sec:intro}

While there is very strong evidence for the existence of dark matter (DM) on the scales of galaxies, galaxy clusters and cosmology \cite{silkreview:2005, bergstromreview:2012}, determining its specific particle properties remains one of the most important open questions in modern physics \cite{snowmasscosmicfrontiers:2014}. There are many elementary particle candidates for the dark matter in the universe (see e.g. \cite{fengreview:2010} for a review of particle DM candidates), but one of the most compelling class of models is based on a striking coincidence: a stable particle with approximately weak-scale mass and couplings (a so-called weakly-interacting massive particle, or WIMP) will, assuming a standard thermal history of the universe, possess a thermal relic density today that roughly corresponds to that of DM. This coincidence is known as ``WIMP miracle''.

Much effort is under way in searching for WIMPs. The most direct probe of WIMPs as DM candidates is the search for scattering of Galactic WIMPs with ordinary matter. Many direct detection experiments have searched for this signal, with no definitive detections as of yet \cite{snowmassdirectdetection:2013, directdetectionreview:2014}. The bounds from such searches have begun to have significant implications for our understanding of viable WIMP DM candidates, ruling out the most na\"\i ve realizations of the WIMP paradigm. Of course, not all DM interactions can be effectively probed by these experiments.

A second technique to search for WIMPs is to attempt to produce them in high-energy particle collisions. Once produced, a WIMP would escape undetected from the experiment, leading to a missing energy signature \cite{snowmassnewparticles:2013}. Searches for missing energy have a very long history, and have been designed for a number of purposes, besides models featuring a stable WIMP.  Recently, a focus on comparing the bounds on interactions of WIMPs with other particles from these searches to indirect and direct detection experiments has emerged.  This exercise started with a parametrization of DM interactions by contact interactions \cite{beltranhooper:2010, shepherdtait:2010, harnikfox:2012}, and has evolved to include various simplified models which feature a collider-accessible mediator particle as well \cite{simplifiedmodelreview:2014}. Many collider analyses are now being designed explicitly with the purpose of searching for DM \cite{colliderplans:2014}.

A third technique, known as indirect detection, searches for the visible particles that result from  annihilations of WIMPs in the Galaxy or in outer systems \cite{snowmassdirectdetection:2013}. These annihilations are rare enough that they do not lead to any significant depletion of the DM density, but they can potentially produce measurable signals. Many experiments have sensitivities good enough to probe annihilation rates that correspond to the rate needed in the early universe in order for the WIMP Miracle to occur \cite{indirectdetectionreview:2014}. We must keep in mind, however, that the kinematics of annihilation at the present time are very different from those needed in the early universe, and thus the annihilation rate may be significantly different as well \cite{crossingsymmetry:2013}.

A variety of target Standard Model (SM) particles can be employed for the purposes of indirect searches. Neutrino telescopes have been used to search for the annihilation of WIMPs captured in the Sun, photons from radio through gamma rays have been used to place bounds on DM annihilation, and cosmic rays can also be employed to search for a WIMP signal. The most promising cosmic rays for such purpose are antimatter particles, as we believe that ordinary astrophysical processes do not produce antimatter in great abundances, and that therefore the only background should be secondary, i.e. arising from cosmic ray interactions with interstellar material. Such secondaries are generally much less abundant and have much less energy than primary cosmic rays, so that WIMP annihilation could potentially yield a detectable primary contribution.

In 2008, the PAMELA experiment precisely measured the positron fraction in the 1.5-100 GeV range \cite{pamelaoriginal:2009}.  The positron fraction was found to sharply increase around 5 GeV in a manner completely inconsistent with secondary sources in essentially all predictive cosmic-ray models \cite{galprop:1998, serpico:2009}.  The Fermi Large Area Telescope (LAT) later confirmed that the positron faction continues to rise in the 100-200 GeV range \cite{fermiep:2012}.  The International Space Station-based experiment Alpha Magnetic Spectrometer (AMS-02) extended measurements to 350 GeV \cite{ams02:2013}.  Their $6.8\times10^6$ electron and positron events confirm the PAMELA excess with high precision.  The slope of the AMS-02 spectrum decreases by an order of magnitude between 20 and 250 GeV and has no measurable fine structure or anisotropy in the arrival direction.  The second AMS-02 data release last fall extended the analysis to 500 GeV, and found that above $\sim 275$ GeV the positron fraction ceases to increase \cite{ams02:2014}.

These experiments imply that there must be another significant source of high energy positrons besides secondary processes. While a Galactic WIMP annihilation origin is still possible \cite{hoopercholis:2013}, nearby pulsars have been proposed as another possible explanation for the positron excess \cite{profumolindenpulsars:2013,yin:2013}.  %Distinguishing between these two possible explanations of the cosmic positron excess may be possible by searching for a small dipole anisotropy in the electron and positron arrival direction, but it appears challenging.

WIMP annihilation into charged states also generically produces photons via loop-level processes or from final-state radiation.  In the latter case, the secondary photons have an energy spectrum peaked towards low energies.  In the former case, however, when DM annihilates at close to zero relative velocity into $\gamma X$, conservation of energy gives a photon with a fixed (monochromatic) energy of
\begin{align}
    E_\gamma &= m_\chi \left( 1 - \frac{m_X^2}{4m_\chi^2} \right), \label{eq:eline}
\end{align}
where $m_\chi$ is the WIMP mass.  A gamma-ray line from these processes could stand out from the continuous astrophysical background for high enough values of $m_\chi$ and be detected by HESS or Fermi LAT.  Since gamma rays propagate in straight lines from their production location, the line of sight of their spatial origin can easily be determined and compared with expectations based on various WIMP models and DM density profiles.  While there was recent excitement about a $130\us{GeV}$ line in the public Fermi LAT data \cite{weniger:2012}, the signal is not significantly present in the Fermi Collaboration's latest analyses based on their Pass 8 software upgrade, and no other spectral lines have been detected \cite{albert130gev:2014}.

\begin{figure}[t]
    \begin{center}
        \begin{fmffile}{tree}
            \begin{fmfgraph*}(50,50)
                \fmfstraight
                \fmfleft{chi,chibar}
                \fmflabel{$\chi$}{chi}
                \fmflabel{$\bar{\chi}$}{chibar}
                \fmfright{f,fbar}
                \fmflabel{$f$}{f}
                \fmflabel{$\bar{f}$}{fbar}
                \fmf{fermion}{chi,v}
                \fmf{fermion}{v,chibar}
                \fmf{fermion}{v,f}
                \fmf{fermion}{fbar,v}
                \fmfblob{25}{v}
            \end{fmfgraph*}
        \end{fmffile}
        \hspace{0.5in}
        \begin{fmffile}{loop}
            \begin{fmfgraph*}(130,50)
                \fmfstraight
                \fmftop{pt1,pt2,pt3,pt4}
                \fmfbottom{pb1,pb2,pb3,pb4}
                \fmfleft{chi,chibar}
                \fmflabel{$\chi$}{chi}
                \fmflabel{$\bar{\chi}$}{chibar}
                \fmfright{gamma2,gamma1}
                \fmf{fermion}{chi,v}
                \fmf{fermion}{v,chibar}
                \fmfblob{25}{v}
                \fmf{fermion}{g1rad,v,g2rad,g1rad}
                \fmf{phantom,tension=5}{pt3,g1rad}
                \fmf{phantom,tension=5}{pb3,g2rad}
                \fmf{photon,tension=1}{g1rad,gamma1}
                \fmf{photon,tension=1}{g2rad,gamma2}
                \fmflabel{$\gamma$}{gamma1}
                \fmflabel{$\gamma$}{gamma2}
            \end{fmfgraph*}
        \end{fmffile}
    \end{center}
	\caption{Feynman diagrams for the processes $\chi\bar{\chi}\to f\bar{f}$ and $\chi\bar{\chi}\to\gamma\gamma$.}
    \label{fig:diagrams}
\end{figure}
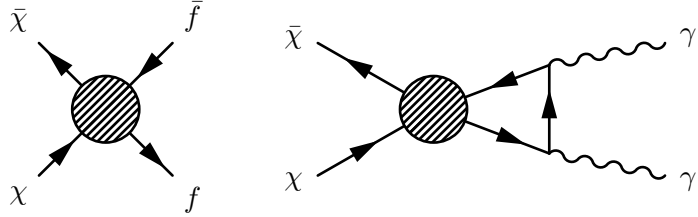

In this paper we study the monochromatic gamma-ray yield stemming from DM models where the DM particle annihilates into a charged lepton-antilepton pair. In particular, we carry out a model-independent study based on contact interactions in the context of an effective field theory (EFT) description of the DM-lepton interaction vertex. For a given contact interaction type, we essentially simply ``close the fermion loop'' as shown in figure \ref{fig:diagrams} and calculate the $\xsecgg$ cross section for a given annihilation cross section into leptons $\xsecff$. As a result, we calculate in a model-independent way the monochromatic gamma-ray yield for e.g. theories that purportedly explain the measured positron excess; vice versa, our results allow to compute $\xsecff$ from a given measured gamma-ray line and model to explain it.

The remainder of this paper is organized as follows:  In section \ref{sec:setup}, we describe the EFTs structure under consideration.  We calculate the ratio $\xsecgg/\xsecff$ in section {\ref{subsec:ratios}}.   In section {\ref{subsec:tchan}}, we comment on the applicability of the EFT assumption in the case of mediators charged under the dark symmetry.  Before concluding, as an example of the utility of our ratio results, we compute the rate of annihilation into monochromatic gamma rays in section {\ref{subsec:gamma_pred}}, assuming a WIMP explanation for the positron excess.

\section{Model Setup}
\label{sec:setup}

We consider a model of WIMP DM where the WIMP is a SM gauge singlet Dirac fermion $\chi$, and we assume that the new particle mediating the interactions between the WIMP and SM fields is heavy enough to be ``integrated out''. The natural way of parametrizing these interactions, then, is to have non-renormalizable contact operators suppressed by a mass scale that is related to the mediator's mass and couplings. In this work we focus on interactions between DM and a SM fermion $f$. The lowest-dimension, and thus least suppressed, operators are therefore dimension six products of DM bilinears and SM bilinears. While we could consider operators containing mixed bilinears as well, these are all equivalent through Fierz transformations to linear combinations of the class we do consider.

The interactions are chosen to be of the form $[\bar{\chi}\Gamma\chi][\bar{f}\Sigma f]$, where $\Gamma$ and $\Sigma$ are matrices in the set $\{1, \gamma^5, \gamma^\mu, \gamma^\mu\gamma^5, \gamma^\mu P_R, \gamma^\mu P_L\}$, with $P_R = \frac{1}{2}(1 + \gamma^5)$ and $P_L = \frac{1}{2} (1 - \gamma^5)$ being the usual helicity projectors.  This is obviously an over-complete basis of operators, as the vector and axial interactions can be constructed out of the chiral interactions or vice versa.  However, due to the chiral nature of the SM gauge interactions the chiral operators may well be of interest, despite their linear dependence on the others. The full list of EFT interactions we consider are shown in Table \ref{tab:efts}.  Our naming scheme is simple.  For scalar operators, $S$ and $P$ stand for scalar ($1$) and pseudoscalar ($\gamma^5$) couplings, respectively.  $V$, $A$, $R$ and $L$ identify vector ($\gamma^\mu$), axial ($\gamma^\mu\gamma^5$), right-handed ($\gamma^\mu P_R$) and left-handed ($\gamma^\mu P_L$) couplings.

\begin{table}[t]
    \centering
    \begin{tabular}{|c|c|c|}
        \hline
        Name & Operator & X\\
        \hline
        SS & $\bar{\chi}\chi \bar{f}f$ & \multirow{4}{*}{$\gamma$}\\
        PS & $\bar{\chi}\gamma^5 \chi \bar{f} f$ & \\
        SP & $\bar{\chi}\chi \bar{f} \gamma^5 f$ & \\
        PP & $\bar{\chi}\gamma^5 \chi \bar{f} \gamma^5 f$ & \\
        \hline
        RR & $\bar{\chi}\gamma^\mu P_R \chi \bar{f} \gamma_\mu P_R f$ & \multirow{4}{*}{$\gamma$}\\
        RL & $\bar{\chi}\gamma^\mu P_R \chi \bar{f} \gamma_\mu P_L f$ & \\
        LR & $\bar{\chi}\gamma^\mu P_L \chi \bar{f} \gamma_\mu P_R f$ & \\
        LL & $\bar{\chi}\gamma^\mu P_L \chi \bar{f} \gamma_\mu P_L f$ & \\
        \hline
    \end{tabular}
    \hspace{0.5in}
    \begin{tabular}{|c|c|c|}
        \hline
        Name & Operator & X\\
        \hline
        VV & $\bar{\chi}\gamma^\mu \chi \bar{f} \gamma_\mu f$ & \multirow{3}{*}{$Z$}\\
        AV & $\bar{\chi}\gamma^\mu \gamma^5 \chi \bar{f} \gamma_\mu f$ & \\
        VA & $\bar{\chi}\gamma^\mu \chi \bar{f} \gamma_\mu \gamma^5 f$ & \\
        \hline
        AA & $\bar{\chi}\gamma^\mu \gamma^5 \chi \bar{f} \gamma_\mu \gamma^5 f$ & $\gamma$\\
        \hline
        RV & $\bar{\chi}\gamma^\mu P_R \chi \bar{f} \gamma_\mu f$ & \multirow{2}{*}{$Z$}\\
        LV & $\bar{\chi}\gamma^\mu P_L \chi \bar{f} \gamma_\mu f$ & \\
        \hline
        RA & $\bar{\chi}\gamma^\mu P_R \chi \bar{f} \gamma_\mu \gamma^5 f$ & \multirow{2}{*}{$\gamma$}\\
        LA & $\bar{\chi}\gamma^\mu P_L \chi \bar{f} \gamma_\mu \gamma^5 f$ & \\
        \hline
        VR & $\bar{\chi}\gamma^\mu \chi \bar{f} \gamma_\mu P_R f$ & \multirow{2}{*}{$Z$}\\
        VL & $\bar{\chi}\gamma^\mu \chi \bar{f} \gamma_\mu P_L f$ & \\
        \hline
        AR & $\bar{\chi}\gamma^\mu \gamma^5 \chi \bar{f} \gamma_\mu P_R f$ & \multirow{2}{*}{$\gamma$}\\
        AL & $\bar{\chi}\gamma^\mu \gamma^5 \chi \bar{f} \gamma_\mu P_L f$ & \\
        \hline
    \end{tabular}
    \caption{Effective operators for WIMP-SM interactions considered in this work.  The $X$ column indicates the second gauge boson in the final state of the $\chi\chi\to\gamma X$ process.}
    \label{tab:efts}
\end{table}

The quantity of interest here is the cross section ratio
\begin{align}
    \rxf{X}{f} \equiv \frac{\xsecgx{X}}{\xsecff},
\end{align}
where $X$ is another SM vector boson.\footnote{Note that the cross sections in this work are calculated at a specific value of the WIMP relative velocity $v$, as opposed to being thermally averaged.}  As discussed before, $\xsecff$ and $\xsecgx{X}$ are computed by evaluating the diagrams shown in figure \ref{fig:diagrams}.  We take $X = \gamma$ for the scalar-type EFTs to maximize photon production.  For operators of the form $[\bar{\chi}\Gamma^\mu\chi][\bar{f}\Sigma_\mu f]$, $X = Z$ unless the WIMP and SM pieces both contain a factor of $\gamma^5$, in which case we also take $X = \gamma$.  See appendix \ref{app:yang} for details behind these choices, which are related to Yang's theorem.

It is important to note that, while we use the language of EFT interactions, any $s$-channel mediator with appropriate couplings will in fact give an identical cross section ratio to our EFT-based approach, as the $s$-channel propagator in both diagrams is identical and the mediator cannot carry the loop momentum. Thus, our resulting ratios are applicable for these theories beyond the regime where the EFT has captured all DM physics of interest. After presenting our EFT results we will discuss the case of $t$-channel mediators in some detail, determining where the EFT analysis holds for $\rxf{X}{f}$ and where it breaks down and the full set of box diagrams must be calculated explicitly.

The AA, AR, AL and AV operators have $p$-wave suppressions at tree level.  This implies that the annihilation rate in the Milky Way halo today is $\sim 10^{-4}$ that in the early universe, at the time when a thermal WIMP would freeze out.  Since the cross section required today to explain the positron excess is instead $\sim 10^3$ times the thermal relic one, this potential explanation of the positron excess requires a non-thermal cosmology, e.g. DM production in moduli decays \cite{Moduli:2013} or a modified expansion history, e.g. due to quintessence \cite{quintessence}.

These suppressions also mean that we cannot consistently take the relative WIMP velocity $v_\chi$ to be zero when computing $\rxf{X}{f}$ for all of our theories, as is often done.  We found that the 3-body processes $\chi\bar{\chi}\to f\bar{f}\gamma$ and $\chi\bar{\chi}\to f\bar{f}Z$ did not significantly alleviate the $p$-wave suppressions, nor the additional helicity suppression found for $\xsecff$ in the AA theory, demonstrating that gamma ray lines are the natural signals for our EFTs.  

We take the velocity for the tree-level process to be $v_{\chi,\tau}/c = 10^{-3}$, a choice motivated by the typical halo velocity dispersion {\cite{dmvelocity:2006}}.  While gamma-ray line searches focus on the Galactic Center, the velocity dispersion in this region is within a factor of two of $v_{\chi,\tau}$ {\cite{dmvelocity:2004}}, so we are justified in using the standard approximation $v_{\chi,\gamma}/c = v_{\chi,\tau} \equiv v_\chi$ to calculate $\xsecgg$.  Accounting for the effect of Sgr A$^*$ on the WIMP velocity dispersion is beyond the scope of this paper.

\section{Results}
\label{sec:results}

\subsection{Cross Section Ratios}
\label{subsec:ratios}
We calculate the ratio $\rxf{X}{\tau}$ in each of the EFTs in Table \ref{tab:efts} using the Mathematica packages FeynArts \cite{feynarts:2001}, FormCalc and LoopTools \cite{formcalclooptools:1999}.  The spin-averaged cross sections are computed with WIMP velocity $v_\chi/c = 10^{-3}$ in the center of mass frame.  We collect our analytic results for the relevant tree-level cross sections in  appendix {\ref{app:tree-analytics}}.  The tree-level and loop-level analytic amplitudes for the SP and AA interactions found in appendix \ref{app:loop-analytics-SP} and \ref{app:loop-analytics-AA} were found to agree with the numerical results.  Notice that while our EFT interactions are non-renormalizable, the one-loop amplitudes are finite.

\begin{figure}[t]
    \begin{center}
        \includegraphics[width=\textwidth]{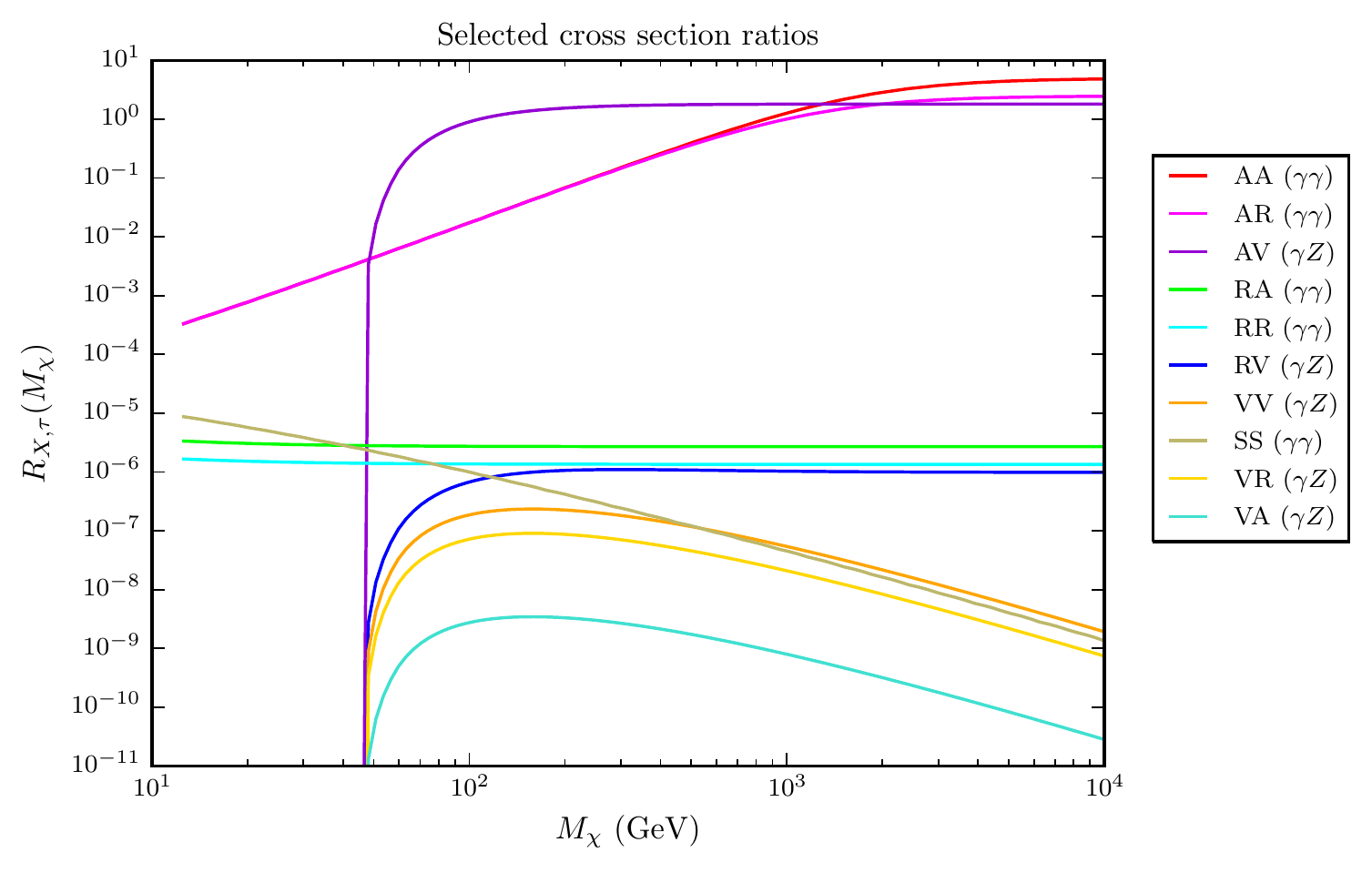}
    \end{center}
	\caption{Cross section ratios for the representative set of the EFTs discussed in section \ref{sec:results}.  The particle name following each EFT label indicates the choice of $X$ that maximizes photon production through the loop diagram.  For theories with $X = Z$, curves begin at the $Z$ production threshold $M_\chi = M_Z / 2 = 45.6\us{GeV}$.}
    \label{fig:ratiosTau}
\end{figure}

The cross section ratios $\rxf{X}{\tau}$ for a representative set of our EFTs are shown in figure \ref{fig:ratiosTau} with WIMP masses ranging from $\sim100\us{GeV}$ to $\sim10\us{TeV}$.  Ratio curves are omitted for some EFTs as they are related simply to others shown in the plot:
\begin{itemize}
    \item
		In the case of the scalar theories, the WIMP and fermion (gamma) parts of the tree-level (loop-level) cross sections decouple.  Since the WIMP components are the same for the tree and loop-level cross sections, the SS and PS results are the same, as are the PS and PP ones.  Moreover, the loop parts of the amplitudes differ by a factor of order 1 for scalar and pseudoscalar coupling to $f$.  We therefore omit the PS, SP and PP results, as they are numerically very similar to the SS one.
    \item
        Averaging over initial WIMP spins allows us to omit LA, LV, LL and LR from our plot, since they give the same results as the RA, RV, RL and RR theories.
    \item
        The AR, AL, RR and RL theories allow annihilations of WIMPs to two photons.  Since photons couplings to $f$ are helicity independent, we also left AL and RL out of our plot.
    \item
		The $\rxf{Z}{\tau}$ ratios for the VR and VL theories differ by a factor $\left( \frac{\sin^2\theta_w}{-1/2+\sin^2\theta_w} \right)^2 \approx 0.6$ since left and right-handed $\tau$s couple differently to the $Z$, so we only show the former.  
\end{itemize}

Most of the EFTs give $\rxf{X}{\tau} \sim 10^{-10} - 10^{-5}$.  However, for AA, AL, AR, $\rxf{X}{\tau}\sim10^{-2} - 10$ and for AV, $\rxf{Z}{\tau}\sim 1$.  While surprising at first that a loop-level to tree-level diagram give such close cross sections, this is easily understood by examining the relevant suppressions.  As mentioned before, the tree-level cross section for the AV theory shown in appendix \ref{app:tree-analytics} is $p$-wave suppressed, but the loop one is not, leading to a large ratio.  Similarly, while the AA, AL and AR theories are $p$-wave and helicity suppressed at tree-level, these suppressions are alleviated in the loop diagrams.  The ratios for these three theories demonstrate the transition between the regime in which the tree-level cross section is dominated by the helicity suppressed term ($M_\chi \lesssim 3\times10^3\us{GeV}$) and the $p$-wave suppressed term ($M_\chi \gtrsim 3\times10^3\us{GeV}$).  In the case $f = \mu$ or $e^-$, the tree-level helicity suppressed terms becomes negligible for these theories, and their ratios are constants equal to their asymptotic values of $1 - 10$.

\subsection{EFT for $t$-channel Mediators}
\label{subsec:tchan}

\begin{figure}[t]
    \begin{center}
        \begin{fmffile}{box}
            \begin{fmfgraph*}(130,50)
                \fmfstraight
                \fmfleft{chi1,chi2}
                \fmfright{gamma1,gamma2}

				\fmf{fermion}{chi1,st1}
				\fmf{dashes,tension=0,label=$\tilde{\tau}$}{st1,st2}
				\fmf{fermion}{st2,chi2}

				\fmf{fermion}{st1,ga1}
				\fmf{fermion,tension=0,label=$\tau$}{ga1,ga2}
				\fmf{fermion}{ga2,st2}

				\fmf{photon}{ga1,gamma1}
				\fmf{photon}{ga2,gamma2}

                \fmflabel{$\chi$}{chi1}
				\fmflabel{$\chi$}{chi2}
                \fmflabel{$\gamma$}{gamma1}
                \fmflabel{$\gamma$}{gamma2}
            \end{fmfgraph*}
        \end{fmffile}
    \end{center}
	\caption{The box diagram contributing to $\xsecgg$ in our theory containing a $t$-channel mediator $\tilde{\tau}$.}
	\label{fig:box}
\end{figure}
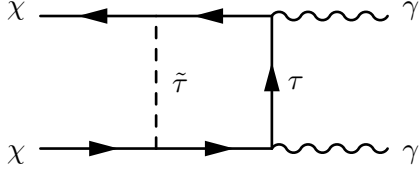

The loop diagrams we consider require care as arbitrarily large loop momenta run through our effective operators.  As mentioned before, if the contact operator comes from an $s$-channel mediator, the mediator is not contained in the loop and the EFT gives the correct result for the ratio of cross sections.  In the case of a charged $t$-channel mediator, we must consider the box diagram shown in figure \ref{fig:box} (along with the crossed ones).  In this case, the mediator's propagator could lead to significant corrections to the cross section from the EFT prediction at high loop momenta.  

To explore this potential issue, we considered a theory in which the WIMP-SM interaction is of the form 
\begin{align}
	\mathcal{L}_\u{int} &= [\bar{\chi} (\lambda_S + \lambda_P \gamma^5) \tau] \tilde{\tau}^+ + (h. c.),
\end{align}
where $\chi$ is the WIMP (taken to be Majorana for simplicity) with mass $m_\chi$ and $\tilde{\tau}^+$ is a charged scalar with mass $m_{\tilde{\tau}}$.  This is a simplified model of a SUSY theory in which one tau slepton is the dominant contributor to DM annihilations and the DM neutralino's gauge eigenstate content is unknown.  We consider the CP-conserving case, which requires that $\lambda_S,\lambda_P\in\mathbb{R}$.  

To lowest order, the effective Lagrangian for this theory will contain a linear combination of four-fermion operators from Table \ref{tab:efts}.  These are found by applying Fierz transformations to the matrix element for $\chi\chi\to\tau^+\tau^-$, giving
\begin{align}
	\mathcal{L}_\u{int} \to \mathcal{L}_\u{eff} %&\supset \frac{|\lambda_S|^2 - |\lambda_P|^2}{4 m_{\tilde{\tau}}^2} \left( \biL{}{} + \biL{\gamma^5}{\gamma^5} \right) + \frac{i \im(\lambda_P \lambda_S^*)}{2 m_{\tilde{\tau}}^2} \left( \biL{\gamma^5}{} + \biL{}{\gamma^5} \right)\\
    %&\hspace{0.5in} - \frac{|\lambda_S|^2 + |\lambda_P|^2}{4 m_{\tilde{\tau}}^2} \biL{\gamma^\mu \gamma^5}{\gamma_\mu \gamma^5} - \frac{\re(\lambda_P \lambda_S^*)}{2 m_{\tilde{\tau}}^2} \biL{\gamma^\mu \gamma^5}{\gamma_\mu}\\
    &= \frac{|\lambda_S|^2 - |\lambda_P|^2}{4 m_{\tilde{\tau}}^2} \left( \biL{}{} + \biL{\gamma^5}{\gamma^5} \right) - \frac{|\lambda_S|^2 + |\lambda_P|^2}{4 m_{\tilde{\tau}}^2} \biL{\gamma^\mu \gamma^5}{\gamma_\mu \gamma^5} \nonumber\\
	&- \frac{\lambda_P \lambda_S}{2 m_{\tilde{\tau}}^2} \biL{\gamma^\mu \gamma^5}{\gamma_\mu}.
	\label{eq:Ltchan}
\end{align}
Note that in the case of a Dirac WIMP, there would be additional terms involving its tensor and vector currents.  In the chiral case $|\lambda_S| = |\lambda_P|$, only the AA and AV terms are present, implying that $\rxf{\gamma}{\tau}$ will be large for this theory.  In what follows we will call the EFT and exact cross sections $\xsecgg^{\u{EFT}}$ and $\xsecgg^{\u{exact}}$.  The latter has been calculated in many papers (see e.g. \cite{oldloopa:1988,oldloopb:1989,oldloopc:1989,neutralinogg:1995,loopCorrection:1997}) in the $v_\chi = 0$ limit.  %Note that $\xsecll{\tau}$ does not vanish in this limit for this theory.  
Here we use the result from \cite{neutralinogg:1995} which is corrected in the footnote on page 8 of \cite{loopCorrection:1997}.

\begin{figure}[t]
    \centering
    \includegraphics{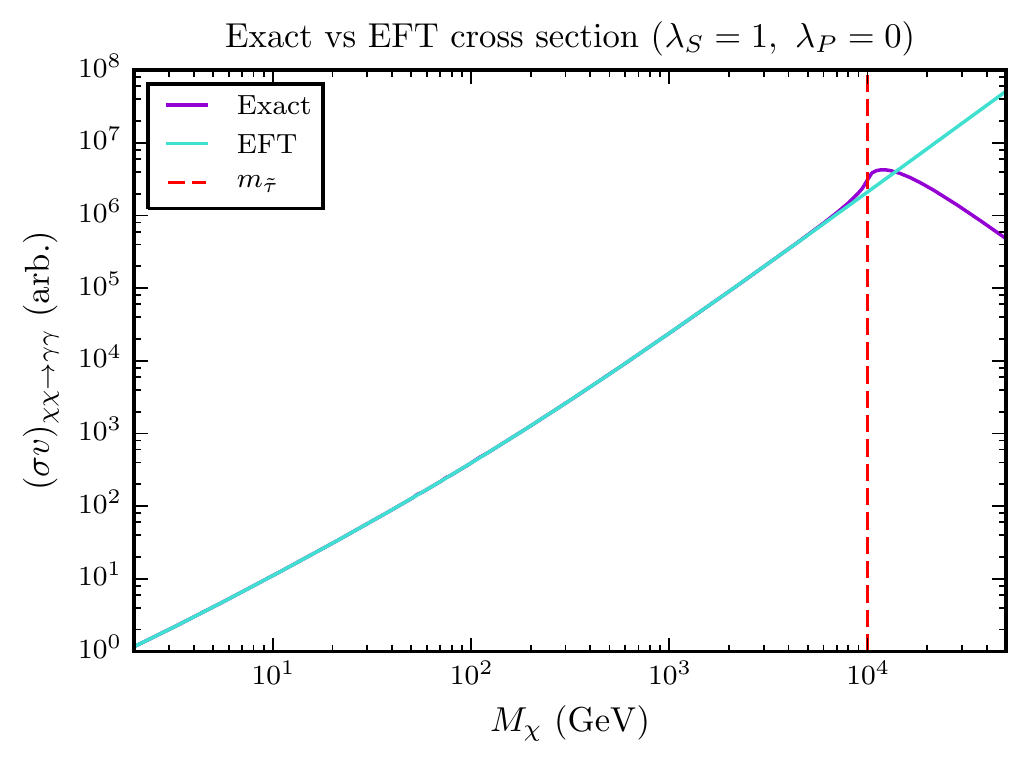}
    \caption{$\xsecgg^{\u{exact}}$ and $\xsecgg^{\u{EFT}}$ plotted as a function of $m_\chi$ for $\lambda_S = 1$ and $\lambda_P = 0$.  The agreement between the two cross sections is better for any other choice of the coupling constants.}
    \label{fig:exact_vs_eft_mchi}
\end{figure}

In figure \ref{fig:exact_vs_eft_mchi} we plot $\xsecgg^{\u{exact}}$ and $\xsecgg^{\u{EFT}}$ as a function of $m_\chi$ for $m_{\tilde{\tau}} = 10\us{TeV}$, $\lambda_S = 0$ and $\lambda_P = 1$.  The cross sections agree to better than $1\%$ all the way up to $m_\chi \approx \frac{1}{2} m_{\tilde{\tau}}$.  Near $m_\chi \approx m_{\tilde{\tau
}}$ the EFT underestimates the cross section when the slepton can be put on shell.  For $m_\chi \gtrsim m_{\tilde{\tau}}$ the EFT result is larger than the exact result since the EFT loop only contains three propagators.  % As expected why?  I forget the reasoning here...
The coupling constants' values have no effect on this agreement.  It is surprising that the EFT is valid for such high values of $m_\chi$, and furthermore that it works so well for our SUSY-like theory, as this is the classic internal bremsstrahlung case in which EFT is usually expected to be inaccurate. 

Finally, we consider whether our calculation of $\rxf{\gamma}{\tau}$ changes in this $t$-channel theory.  The cross section $\xsecff$ gets a new contribution from the internal bremsstrahlung diagram for $\chi\bar{\chi} \to f \bar{f} \gamma$.  This diagram only contributes significantly to $\xsecff$ if the helicity suppression in the lowest-order process $\chi\bar{\chi}\to f\bar{f}$ is very large.  Since we are interested in the final state $f = \tau$, this is not the case \cite{oldloopb:1989}, and we can safely ignore the internal bremsstrahlung diagram for the range of WIMP masses considered above; for the largest WIMP masses considered the helicity suppression is comparable to the one-loop suppression, leading to a change of only a factor of a few in the ratio.  Our type of EFT analysis of $\rxf{\gamma}{\tau}$ from above thus applies for $m_\chi \lesssim \frac{1}{2} m_{\tilde{\tau}}$, although of course having to consider a linear combination of effective operators in eq. \eqref{eq:Ltchan} adds additional complexity.  For a lighter final state fermion or much heavier WIMP, a more careful analysis of $\xsecff$ including the internal bremsstrahlung diagram would have to be performed, but note that the WIMP mass for a simplified theory of this type has been shown to have an upper bound comparable to the masses considered here \cite{cahill-rowleyhedri:2015}.

\subsection{Predicting the Gamma Ray Flux}
\label{subsec:gamma_pred}

\begin{figure}[t]
    \begin{center}
        \includegraphics[width=\textwidth]{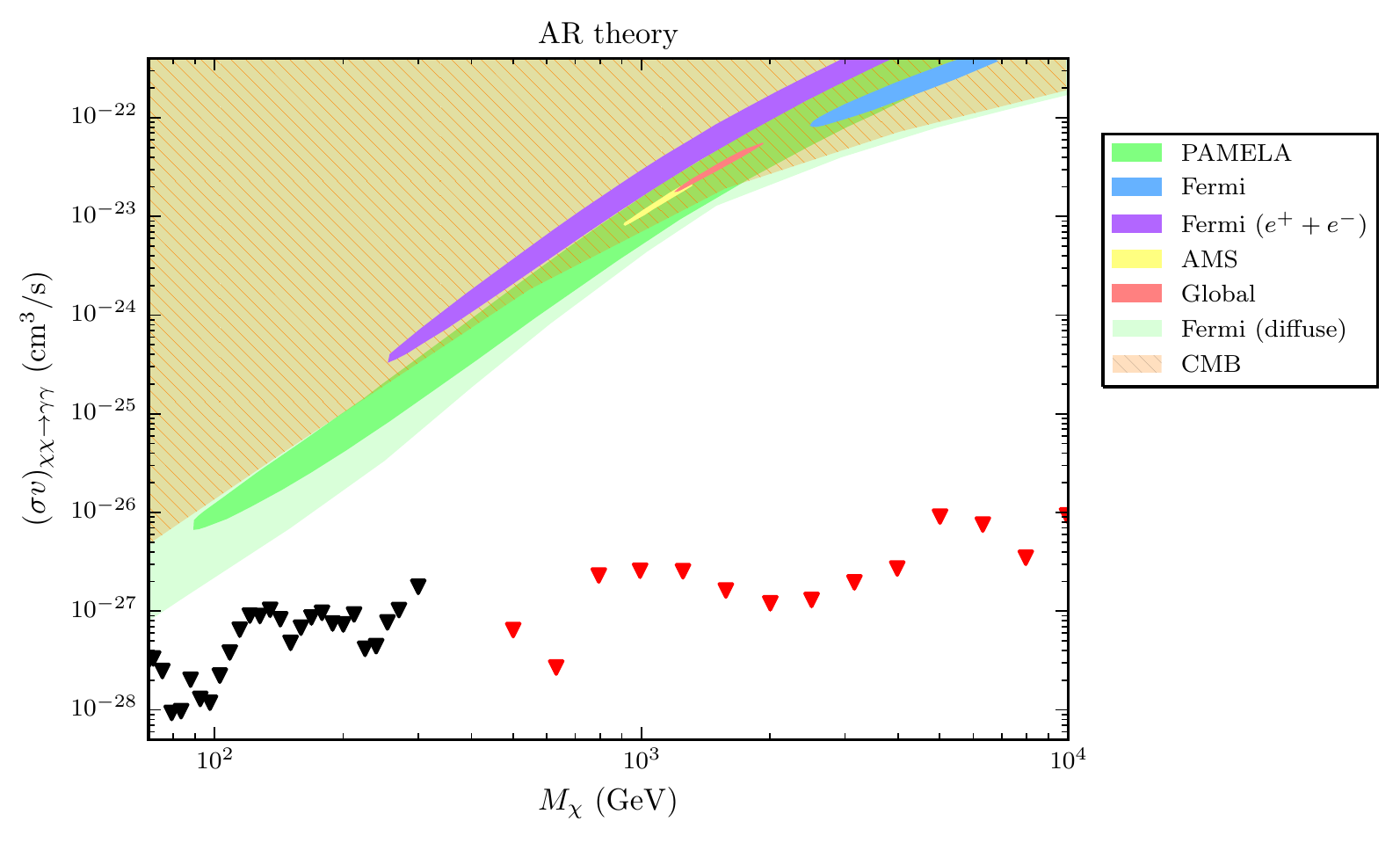}
    \end{center}
	\caption{$(M_\chi,\xsecgg)$ contours for the AR theory predicted using the calculated ratio $\rxf{\gamma}{\tau}$ and preferred regions at $99\%$ CL obtained from positron data in \cite{amsFits:2013a}, for $v_\chi/c=10^{-3}$.}
    \label{fig:gamma_from_tau}
\end{figure}

\begin{figure}[t]
    \begin{center}
        \includegraphics[width=\textwidth]{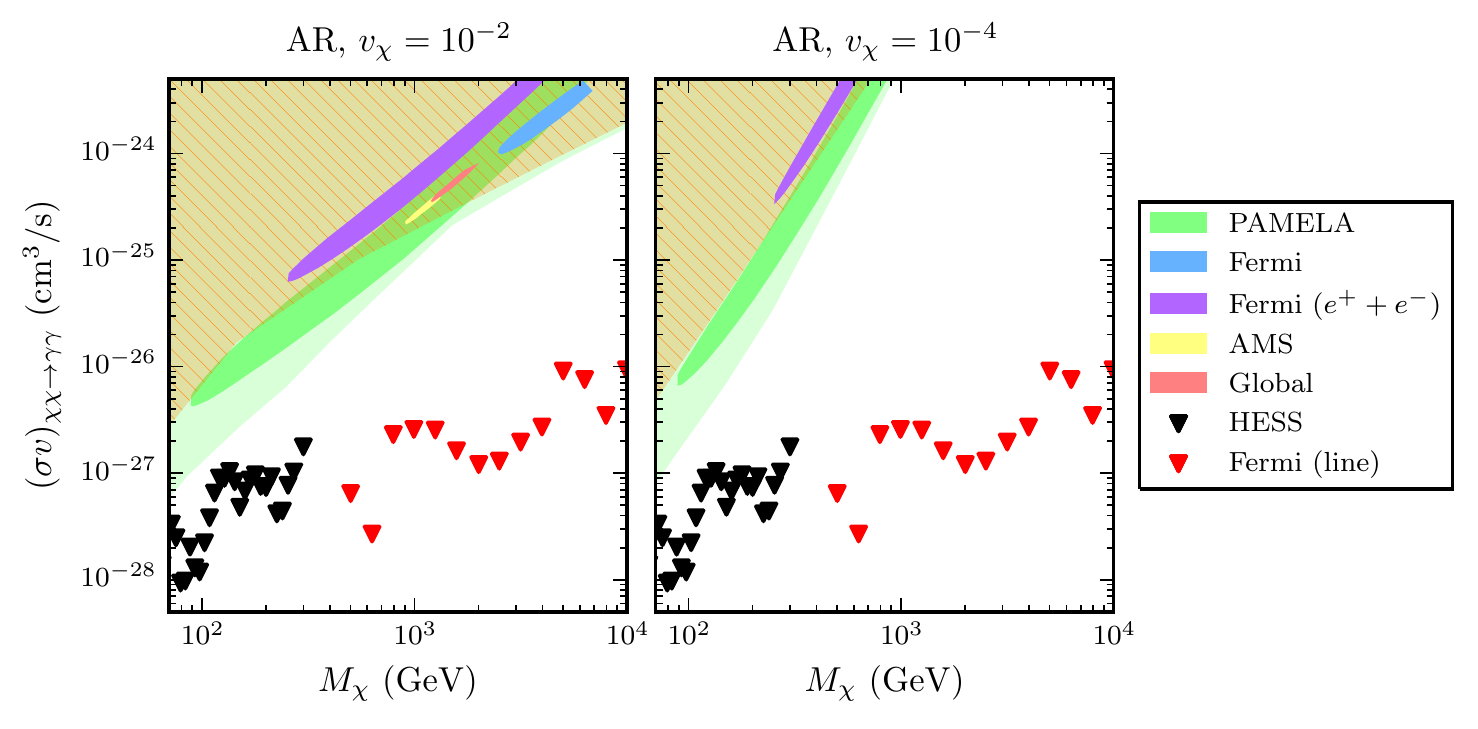}
    \end{center}
    \caption{Same as in figure \ref{fig:gamma_from_tau}, but with $v_\chi$ increased (decreased) by an order of magnitude in the left (right) plot.}
    \label{fig:gamma_from_tau_vel}
\end{figure}

As an example of the utility of our results, suppose now we postulate a WIMP annihilation explanation for the rising positron fraction.  Analyses of the first release of AMS-02 data found that TeV-scale WIMPs annihilating to $\tau\tau$ provide the best explanation of the positron excess in terms of DM annihilating to SM particles \cite{amsFits:2013a,amsFits:2013b,amsFits:2013c}.  This has not changed with AMS-02's latest data release \cite{amsfitsupdate:2014}.  To make quantitative predictions for the gamma-ray line yield, we use for reference the allowed regions in the $(m_\chi, \xsecll{\tau})$ plane from \cite{amsFits:2013a}.  Our calculated ratios enable us to predict the gamma-ray monochromatic line signal induced through the loop diagram for a given contact interaction.  

By rescaling the $99\%$ CL allowed regions in the $(M_\chi, \xsecll{\tau})$ plane given in \cite{amsFits:2013a} by $\rxf{X}{\tau}$ and putting in a factor of $1/2$ for theories with $X = Z$, we calculated the regions in the $(M_\chi,\xsecgg)$ plane preferred by a WIMP annihilation explanation to the positron fraction anomaly.  Figure \ref{fig:gamma_from_tau} shows the resulting regions for the AR EFT along with recent upper limits on $\xsecgg$ from HESS \cite{hesslinesearch:2013} and Fermi LAT \cite{fermilinesearch:2013} line searches.  The predicted allowed regions for AA, AL and AV lie above those for AR while the regions for all other theories fall well below current observational constraints, by at least 3 orders of magnitude.  HESS-II, GAMMA-400 and CTA line searches are only expected to tighten these constraints by an order of magnitude \cite{futurelinesearches:2012}, and so will not be able to probe line signals from AMS-fitting points in these theories.  

For comparison, we also show the latest constraints on $\xsecll{\tau}$ based on CMB data from Planck and other experiments \cite{cmb:2014} as well as Fermi's observations of diffuse gamma-ray emission in the Milky Way {\cite{fermidiffuse:2012}}, which are in tension with WIMP explanations of the positron fraction.  Note that the rescaled CMB constraints are quite conservative at high $M_\chi$.  In this part of parameter space, the $p$-wave term dominates the tree-level cross section, which means the limits are a factor $v_{\chi,\tau} / v_{\chi,\u{recom}} \gg 1$ larger, where $v_{\chi,\u{recom}}$ is the characteristic relative WIMP velocity at recombination.

Figure \ref{fig:gamma_from_tau_vel} shows the velocity dependence of the results for the AR theory.  The differences between the two panels stem from changes in the relative strengths of the helicity and $p$-wave suppressed parts of the tree-level cross section.  While increasing or decreasing $v_\chi$ by an order of magnitude changes the shape of the limit regions, it does not affect whether they fall above or below the line search constraints for any of our theories.  We therefore find that AA, AR, AL and AV contact interactions cannot explain the positron excess as they would violate bounds from gamma-ray line searches, in addition to being in tension with Fermi diffuse and CMB constraints.  All other contact interactions remain viable.

\section{Discussion and Conclusions}
\label{sec:conclusion}

In this study we have calculated the cross section for Dirac WIMPs annihilating to monochromatic gamma rays stemming from an annihilation mode to a charged lepton pair. To relate $\xsecgg$ and $\xsecff$, we used an over-complete set of dimension 6 operators.  The ratio $\rxf{X}{\tau}$ varies by many orders of magnitude depending on the theory under consideration, since helicity and $p$-wave suppressions in the tree-level cross section are sometimes alleviated at one loop.  We also investigated the validity of the EFT description for a specific SUSY-inspired case ($t$-channel stau-mediated annihilation into tau pair) and demonstrated that the EFT accurately describes theories where a charged boson mediates SM-WIMP interactions.  Since the EFT gives the same value for $\rxf{X}{\tau}$ as an $s$-channel mediator, this shows that our analyses hold for a wide range of theories whose UV completion involves a mediator heavier than the WIMP.

We used our cross section ratios to predict the gamma ray signal given an annihilating WIMP explanation for the positron excess.  Describing the positron excess using contact operators involving an axial WIMP bilinear (namely the AA, AR, AL and AV operators) gives a monochromatic gamma-ray line orders of magnitude brighter than current observational limits from Fermi LAT and HESS.  Our approach is complementary to constraints on diffuse gamma ray emission and CMB measurements, and will prove useful in directing gamma ray searches should a signal from WIMPs annihilating to leptons be detected in the future.

\section*{Acknowledgements}
This research is partially supported by the US Department of Energy, Contract DE-FG02-04ER41286.  A. C. thanks Michael Wagman, Jonathan Cornell and Patrick Draper for useful discussions.

\bibliography{positrons,gamma_lines,neutralino_loops,yangs_theorem,general_dm,feyn,eft,simplified_models}

\providecommand{\href}[2]{#2}\begingroup\raggedright\begin{thebibliography}{10}

\bibitem{silkreview:2005}
G.~Bertone, D.~Hooper, and J.~Silk, {\it Particle dark matter: Evidence,
  candidates and constraints},  {\em Physics Reports} {\bf 405} (2005),
  no.~5–6 279 -- 390, [\href{http://arxiv.org/abs/hep-ph/040}{{\tt
  hep-ph/040}}].

\bibitem{bergstromreview:2012}
L.~Bergstr{\"o}m, {\it Dark matter evidence, particle physics candidates and
  detection methods},  {\em Annalen der Physik} {\bf 524} (2012), no.~9-10
  479--496, [\href{http://arxiv.org/abs/astro-ph/1}{{\tt astro-ph/1}}].

\bibitem{snowmasscosmicfrontiers:2014}
J.~L. Feng, S.~Ritz, J.~J. Beatty, J.~Buckley, D.~F. Cowen, P.~Cushman,
  S.~Dodelson, C.~Galbiati, K.~Honscheid, D.~Hooper, M.~Kaplinghat, A.~Kusenko,
  K.~Matchev, D.~McKinsey, A.~E. Nelson, A.~Olinto, S.~Profumo, H.~Robertson,
  L.~Rosenberg, G.~Sinnis, and T.~M.~P. Tait, {\it Planning the future of u.s.
  particle physics (snowmass 2013): Chapter 4: Cosmic frontier},  2014.

\bibitem{fengreview:2010}
J.~L. Feng, {\it Dark matter candidates from particle physics and methods of
  detection},  {\em Annual Review of Astronomy and Astrophysics} {\bf 48}
  (2010), no.~1 495--545, [\href{http://arxiv.org/abs/astro-ph/1}{{\tt
  astro-ph/1}}].

\bibitem{snowmassdirectdetection:2013}
P.~Cushman, C.~Galbiati, D.~N. McKinsey, H.~Robertson, T.~M.~P. Tait, D.~Bauer,
  A.~Borgland, B.~Cabrera, F.~Calaprice, J.~Cooley, T.~Empl, R.~Essig,
  E.~Figueroa-Feliciano, R.~Gaitskell, S.~Golwala, J.~Hall, R.~Hill, A.~Hime,
  E.~Hoppe, L.~Hsu, E.~Hungerford, R.~Jacobsen, M.~Kelsey, R.~F. Lang, W.~H.
  Lippincott, B.~Loer, S.~Luitz, V.~Mandic, J.~Mardon, J.~Maricic, R.~Maruyama,
  R.~Mahapatra, H.~Nelson, J.~Orrell, K.~Palladino, E.~Pantic, R.~Partridge,
  A.~Ryd, T.~Saab, B.~Sadoulet, R.~Schnee, W.~Shepherd, A.~Sonnenschein,
  P.~Sorensen, M.~Szydagis, T.~Volansky, M.~Witherell, D.~Wright, and K.~Zurek,
  {\it Snowmass cf1 summary: Wimp dark matter direct detection},  2013.

\bibitem{directdetectionreview:2014}
L.~Baudis, {\it Wimp dark matter direct-detection searches in noble gases},
  {\em Physics of the Dark Universe} {\bf 4} (2014), no.~0 50 -- 59,
  [\href{http://arxiv.org/abs/astro-ph/1}{{\tt astro-ph/1}}]. DARK TAUP2013.

\bibitem{snowmassnewparticles:2013}
Y.~Gershtein, M.~Luty, M.~Narain, L.~T. Wang, D.~Whiteson, K.~Agashe,
  L.~Apanasevich, G.~Artoni, A.~Avetisyan, H.~Baer, C.~Bartels, M.~Bauer,
  D.~Berge, M.~Berggren, S.~Bhattacharya, K.~Black, T.~Bose, J.~Brau, R.~Brock,
  E.~Brownson, M.~Cahill-Rowley, A.~Cakir, A.~Chaus, T.~Cohen, B.~Coleppa,
  R.~Cotta, N.~Craig, K.~Dienes, B.~Dobrescu, D.~Duggan, R.~Essig, J.~Evans,
  A.~Drlica-Wagner, S.~Funk, J.~George, F.~Goertz, T.~Golling, T.~Han, A.~Haas,
  M.~Hance, D.~Hayden, U.~Heintz, A.~Henrichs, J.~Hewett, J.~Hirschauer,
  K.~Howe, A.~Ismail, K.~Kaadze, Y.~Kats, F.~Kling, D.~Kolchmeyer,
  D.~Kr``ucker, K.~C. Kong, A.~Kumar, G.~Kribs, P.~Langacker, A.~Lath, S.~J.
  Lee, J.~List, T.~Lin, L.~Linssen, T.~Liu, Z.~Liu, A.~Lobanov, J.~Loyal,
  A.~Martin, I.~Melzer-Pellmann, M.~M. Nojiri, S.~Padhi, N.~Parashar,
  B.~Penning, M.~Perelstein, M.~Peskin, A.~Pierce, W.~Porod, C.~Potter,
  T.~Rizzo, G.~Sciolla, J.~S. III, S.~Su, T.~M.~P. Tait, T.~Tanabe, B.~Thomas,
  S.~Thomas, S.~Upadhyay, N.~Varelas, E.~Varnes, L.~Vecchi, A.~Venturini,
  B.~Vormwald, J.~Wacker, M.~Walker, M.~Wood, F.~Yu, and N.~Zhou, {\it New
  particles working group report of the snowmass 2013 community summer study},
  2013.

\bibitem{beltranhooper:2010}
M.~Beltr{\'a}n, D.~Hooper, E.~Kolb, Z.~Krusberg, and T.~Tait, {\it Maverick
  dark matter at colliders},  {\em Journal of High Energy Physics} {\bf 2010}
  (2010), no.~9 [\href{http://arxiv.org/abs/hep-ph/100}{{\tt hep-ph/100}}].

\bibitem{shepherdtait:2010}
J.~Goodman, M.~Ibe, A.~Rajaraman, W.~Shepherd, T.~M.~P. Tait, and H.-B. Yu,
  {\it Constraints on dark matter from colliders},  {\em Phys. Rev. D} {\bf 82}
  (Dec, 2010) 116010, [\href{http://arxiv.org/abs/hep-ph/100}{{\tt
  hep-ph/100}}].

\bibitem{harnikfox:2012}
P.~J. Fox, R.~Harnik, J.~Kopp, and Y.~Tsai, {\it Missing energy signatures of
  dark matter at the lhc},  {\em Phys. Rev. D} {\bf 85} (Mar, 2012) 056011,
  [\href{http://arxiv.org/abs/hep-ph/110}{{\tt hep-ph/110}}].

\bibitem{simplifiedmodelreview:2014}
J.~Abdallah, A.~Ashkenazi, A.~Boveia, G.~Busoni, A.~De~Simone, et~al., {\it
  Simplified models for dark matter and missing energy searches at the lhc},
  \href{http://arxiv.org/abs/hep-ph/140}{{\tt hep-ph/140}}.

\bibitem{colliderplans:2014}
A.~Askew, S.~Chauhan, B.~Penning, W.~Shepherd, and M.~Tripathi, {\it Searching
  for dark matter at hadron colliders},  {\em Int. J. Mod. Phys.} {\bf A29}
  (2014) 1430041, [\href{http://arxiv.org/abs/hep-ph/140}{{\tt hep-ph/140}}].

\bibitem{indirectdetectionreview:2014}
S.~Funk, {\it Indirect detection of dark matter with γ rays},  {\em
  Proceedings of the National Academy of Sciences} (2014)
  [\href{http://arxiv.org/abs/astro-ph/1}{{\tt astro-ph/1}}].

\bibitem{crossingsymmetry:2013}
S.~Profumo, W.~Shepherd, and T.~M.~P. Tait, {\it Pitfalls of dark matter
  crossing symmetries},  {\em Phys. Rev. D} {\bf 88} (Sep, 2013) 056018,
  [\href{http://arxiv.org/abs/hep-ph/130}{{\tt hep-ph/130}}].

\bibitem{pamelaoriginal:2009}
O.~Adriani, G.~C. Barbarino, G.~A. Bazilevskaya, R.~Bellotti, M.~Boezio, E.~A.
  Bogomolov, L.~Bonechi, M.~Bongi, V.~Bonvicini, S.~Bottai, A.~Bruno,
  F.~Cafagna, D.~Campana, P.~Carlson, M.~Casolino, G.~Castellini, M.~P.
  De~Pascale, G.~De~Rosa, N.~De~Simone, V.~Di~Felice, A.~M. Galper,
  L.~Grishantseva, P.~Hofverberg, S.~V. Koldashov, S.~Y. Krutkov, A.~N.
  Kvashnin, A.~Leonov, V.~Malvezzi, L.~Marcelli, W.~Menn, V.~V. Mikhailov,
  E.~Mocchiutti, S.~Orsi, G.~Osteria, P.~Papini, M.~Pearce, P.~Picozza,
  M.~Ricci, S.~B. Ricciarini, M.~Simon, R.~Sparvoli, P.~Spillantini, Y.~I.
  Stozhkov, A.~Vacchi, E.~Vannuccini, G.~Vasilyev, S.~A. Voronov, Y.~T. Yurkin,
  G.~Zampa, N.~Zampa, and V.~G. Zverev, {\it An anomalous positron abundance in
  cosmic rays with energies 1.5--100 gev},  {\em Nature} {\bf 458} (April,
  2009) 607--609, [\href{http://arxiv.org/abs/astro-ph/0}{{\tt astro-ph/0}}].

\bibitem{galprop:1998}
A.~W. Strong and I.~V. Moskalenko, {\it Propagation of cosmic-ray nucleons in
  the galaxy},  {\em The Astrophysical Journal} {\bf 509} (1998), no.~1 212,
  [\href{http://arxiv.org/abs/astro-ph/9}{{\tt astro-ph/9}}].

\bibitem{serpico:2009}
P.~D. Serpico, {\it Possible causes of a rise with energy of the cosmic ray
  positron fraction},  {\em Phys. Rev. D} {\bf 79} (Jan, 2009) 021302,
  [\href{http://arxiv.org/abs/hep-ph/081}{{\tt hep-ph/081}}].

\bibitem{fermiep:2012}
M.~Ackermann, M.~Ajello, A.~Allafort, W.~B. Atwood, L.~Baldini, G.~Barbiellini,
  D.~Bastieri, K.~Bechtol, R.~Bellazzini, B.~Berenji, R.~D. Blandford, E.~D.
  Bloom, E.~Bonamente, A.~W. Borgland, A.~Bouvier, J.~Bregeon, M.~Brigida,
  P.~Bruel, R.~Buehler, S.~Buson, G.~A. Caliandro, R.~A. Cameron, P.~A.
  Caraveo, J.~M. Casandjian, C.~Cecchi, E.~Charles, A.~Chekhtman, C.~C. Cheung,
  J.~Chiang, S.~Ciprini, R.~Claus, J.~Cohen-Tanugi, J.~Conrad, S.~Cutini,
  A.~de~Angelis, F.~de~Palma, C.~D. Dermer, S.~W. Digel, E.~do~Couto~e Silva,
  P.~S. Drell, A.~Drlica-Wagner, C.~Favuzzi, S.~J. Fegan, E.~C. Ferrara, W.~B.
  Focke, P.~Fortin, Y.~Fukazawa, S.~Funk, P.~Fusco, F.~Gargano, D.~Gasparrini,
  S.~Germani, N.~Giglietto, P.~Giommi, F.~Giordano, M.~Giroletti, T.~Glanzman,
  G.~Godfrey, I.~A. Grenier, J.~E. Grove, S.~Guiriec, M.~Gustafsson,
  D.~Hadasch, A.~K. Harding, M.~Hayashida, R.~E. Hughes, G.~J\'ohannesson,
  A.~S. Johnson, T.~Kamae, H.~Katagiri, J.~Kataoka, J.~Kn\"odlseder, M.~Kuss,
  J.~Lande, L.~Latronico, M.~Lemoine-Goumard, M.~Llena~Garde, F.~Longo,
  F.~Loparco, M.~N. Lovellette, P.~Lubrano, G.~M. Madejski, M.~N. Mazziotta,
  J.~E. McEnery, P.~F. Michelson, W.~Mitthumsiri, T.~Mizuno, A.~A. Moiseev,
  C.~Monte, M.~E. Monzani, A.~Morselli, I.~V. Moskalenko, S.~Murgia,
  T.~Nakamori, P.~L. Nolan, J.~P. Norris, E.~Nuss, M.~Ohno, T.~Ohsugi,
  A.~Okumura, N.~Omodei, E.~Orlando, J.~F. Ormes, M.~Ozaki, D.~Paneque,
  D.~Parent, M.~Pesce-Rollins, M.~Pierbattista, F.~Piron, G.~Pivato, T.~A.
  Porter, S.~Rain\`o, R.~Rando, M.~Razzano, S.~Razzaque, A.~Reimer, O.~Reimer,
  T.~Reposeur, S.~Ritz, R.~W. Romani, M.~Roth, H.~F.-W. Sadrozinski, C.~Sbarra,
  T.~L. Schalk, C.~Sgr\`o, E.~J. Siskind, G.~Spandre, P.~Spinelli, A.~W.
  Strong, H.~Takahashi, T.~Takahashi, T.~Tanaka, J.~G. Thayer, J.~B. Thayer,
  L.~Tibaldo, M.~Tinivella, D.~F. Torres, G.~Tosti, E.~Troja, Y.~Uchiyama,
  T.~L. Usher, J.~Vandenbroucke, V.~Vasileiou, G.~Vianello, V.~Vitale, A.~P.
  Waite, B.~L. Winer, K.~S. Wood, M.~Wood, Z.~Yang, and S.~Zimmer, {\it
  Measurement of separate cosmic-ray electron and positron spectra with the
  fermi large area telescope},  {\em Phys. Rev. Lett.} {\bf 108} (Jan, 2012)
  011103, [\href{http://arxiv.org/abs/astro-ph/1}{{\tt astro-ph/1}}].

\bibitem{ams02:2013}
M.~Aguilar, G.~Alberti, B.~Alpat, A.~Alvino, G.~Ambrosi, K.~Andeen,
  H.~Anderhub, L.~Arruda, P.~Azzarello, A.~Bachlechner, F.~Barao, B.~Baret,
  A.~Barrau, L.~Barrin, A.~Bartoloni, L.~Basara, A.~Basili, L.~Batalha,
  J.~Bates, R.~Battiston, J.~Bazo, R.~Becker, U.~Becker, M.~Behlmann,
  B.~Beischer, J.~Berdugo, P.~Berges, B.~Bertucci, G.~Bigongiari, A.~Biland,
  V.~Bindi, S.~Bizzaglia, G.~Boella, W.~de~Boer, K.~Bollweg, J.~Bolmont,
  B.~Borgia, S.~Borsini, M.~J. Boschini, G.~Boudoul, M.~Bourquin, P.~Brun,
  M.~Bu\'enerd, J.~Burger, W.~Burger, F.~Cadoux, X.~D. Cai, M.~Capell,
  D.~Casadei, J.~Casaus, V.~Cascioli, G.~Castellini, I.~Cernuda, F.~Cervelli,
  M.~J. Chae, Y.~H. Chang, A.~I. Chen, C.~R. Chen, H.~Chen, G.~M. Cheng, H.~S.
  Chen, L.~Cheng, N.~Chernoplyiokov, A.~Chikanian, E.~Choumilov, V.~Choutko,
  C.~H. Chung, C.~Clark, R.~Clavero, G.~Coignet, V.~Commichau, C.~Consolandi,
  A.~Contin, C.~Corti, M.~T. Costado~Dios, B.~Coste, D.~Crespo, Z.~Cui, M.~Dai,
  C.~Delgado, S.~Della~Torre, B.~Demirkoz, P.~Dennett, L.~Derome, S.~Di~Falco,
  X.~H. Diao, A.~Diago, L.~Djambazov, C.~D\'iaz, P.~von Doetinchem, W.~J. Du,
  J.~M. Dubois, R.~Duperay, M.~Duranti, D.~D'Urso, A.~Egorov, A.~Eline, F.~J.
  Eppling, T.~Eronen, J.~van Es, H.~Esser, A.~Falvard, E.~Fiandrini,
  A.~Fiasson, E.~Finch, P.~Fisher, K.~Flood, R.~Foglio, M.~Fohey, S.~Fopp,
  N.~Fouque, Y.~Galaktionov, M.~Gallilee, L.~Gallin-Martel, G.~Gallucci,
  B.~Garc\'ia, J.~Garc\'ia, R.~Garc\'ia-L\'opez, L.~Garc\'ia-Tabares,
  C.~Gargiulo, H.~Gast, I.~Gebauer, S.~Gentile, M.~Gervasi, W.~Gillard,
  F.~Giovacchini, L.~Girard, P.~Goglov, J.~Gong, C.~Goy-Henningsen, D.~Grandi,
  M.~Graziani, A.~Grechko, A.~Gross, I.~Guerri, C.~de~la Gu\'ia, K.~H. Guo,
  M.~Habiby, S.~Haino, F.~Hauler, Z.~H. He, M.~Heil, J.~Heilig, R.~Hermel,
  H.~Hofer, Z.~C. Huang, W.~Hungerford, M.~Incagli, M.~Ionica, A.~Jacholkowska,
  W.~Y. Jang, H.~Jinchi, M.~Jongmanns, L.~Journet, L.~Jungermann, W.~Karpinski,
  G.~N. Kim, K.~S. Kim, T.~Kirn, R.~Kossakowski, A.~Koulemzine, O.~Kounina,
  A.~Kounine, V.~Koutsenko, M.~S. Krafczyk, E.~Laudi, G.~Laurenti,
  C.~Lauritzen, A.~Lebedev, M.~W. Lee, S.~C. Lee, C.~Leluc, H.~Le\'on~Vargas,
  V.~Lepareur, J.~Q. Li, Q.~Li, T.~X. Li, W.~Li, Z.~H. Li, P.~Lipari, C.~H.
  Lin, D.~Liu, H.~Liu, T.~Lomtadze, Y.~S. Lu, S.~Lucidi, K.~L\"ubelsmeyer,
  J.~Z. Luo, W.~Lustermann, S.~Lv, J.~Madsen, R.~Majka, A.~Malinin,
  C.~Ma\~n\'a, J.~Mar\'in, T.~Martin, G.~Mart\'inez, F.~Masciocchi, N.~Masi,
  D.~Maurin, A.~McInturff, P.~McIntyre, A.~Menchaca-Rocha, Q.~Meng,
  M.~Menichelli, I.~Mereu, M.~Millinger, D.~C. Mo, M.~Molina, P.~Mott,
  A.~Mujunen, S.~Natale, P.~Nemeth, J.~Q. Ni, N.~Nikonov, F.~Nozzoli, P.~Nunes,
  A.~Obermeier, S.~Oh, A.~Oliva, F.~Palmonari, C.~Palomares, M.~Paniccia,
  A.~Papi, W.~H. Park, M.~Pauluzzi, F.~Pauss, A.~Pauw, E.~Pedreschi,
  S.~Pensotti, R.~Pereira, E.~Perrin, G.~Pessina, G.~Pierschel, F.~Pilo,
  A.~Piluso, C.~Pizzolotto, V.~Plyaskin, J.~Pochon, M.~Pohl, V.~Poireau,
  S.~Porter, J.~Pouxe, A.~Putze, L.~Quadrani, X.~N. Qi, P.~G. Rancoita,
  D.~Rapin, Z.~L. Ren, J.~S. Ricol, E.~Riihonen, I.~Rodr\'iguez, U.~Roeser,
  S.~Rosier-Lees, L.~Rossi, A.~Rozhkov, D.~Rozza, A.~Sabellek, R.~Sagdeev,
  J.~Sandweiss, B.~Santos, P.~Saouter, M.~Sarchioni, S.~Schael, D.~Schinzel,
  M.~Schmanau, G.~Schwering, A.~Schulz~von Dratzig, G.~Scolieri, E.~S. Seo,
  B.~S. Shan, J.~Y. Shi, Y.~M. Shi, T.~Siedenburg, R.~Siedling, D.~Son,
  F.~Spada, F.~Spinella, M.~Steuer, K.~Stiff, W.~Sun, W.~H. Sun, X.~H. Sun,
  M.~Tacconi, C.~P. Tang, X.~W. Tang, Z.~C. Tang, L.~Tao, J.~Tassan-Viol,
  S.~C.~C. Ting, S.~M. Ting, C.~Titus, N.~Tomassetti, F.~Toral, J.~Torsti,
  J.~R. Tsai, J.~C. Tutt, J.~Ulbricht, T.~Urban, V.~Vagelli, E.~Valente,
  C.~Vannini, E.~Valtonen, M.~Vargas~Trevino, S.~Vaurynovich, M.~Vecchi,
  M.~Vergain, B.~Verlaat, C.~Vescovi, J.~P. Vialle, G.~Viertel, G.~Volpini,
  D.~Wang, N.~H. Wang, Q.~L. Wang, R.~S. Wang, X.~Wang, Z.~X. Wang,
  W.~Wallraff, Z.~L. Weng, M.~Willenbrock, M.~Wlochal, H.~Wu, K.~Y. Wu, Z.~S.
  Wu, W.~J. Xiao, S.~Xie, R.~Q. Xiong, G.~M. Xin, N.~S. Xu, W.~Xu, Q.~Yan,
  J.~Yang, M.~Yang, Q.~H. Ye, H.~Yi, Y.~J. Yu, Z.~Q. Yu, S.~Zeissler, J.~G.
  Zhang, Z.~Zhang, M.~M. Zhang, Z.~M. Zheng, H.~L. Zhuang, V.~Zhukov,
  A.~Zichichi, P.~Zuccon, and C.~Zurbach, {\it First result from the alpha
  magnetic spectrometer on the international space station: Precision
  measurement of the positron fraction in primary cosmic rays of 0.5--350 gev},
   {\em Phys. Rev. Lett.} {\bf 110} (Apr, 2013) 141102.

\bibitem{ams02:2014}
{\bf (AMS Collaboration)} Collaboration, L.~Accardo, M.~Aguilar, D.~Aisa,
  B.~Alpat, A.~Alvino, G.~Ambrosi, K.~Andeen, L.~Arruda, N.~Attig,
  P.~Azzarello, A.~Bachlechner, F.~Barao, A.~Barrau, L.~Barrin, A.~Bartoloni,
  L.~Basara, M.~Battarbee, R.~Battiston, J.~Bazo, U.~Becker, M.~Behlmann, and
  B.~Beischer, {\it High statistics measurement of the positron fraction in
  primary cosmic rays of 0.5--500 gev with the alpha magnetic spectrometer on
  the international space station},  {\em Phys. Rev. Lett.} {\bf 113} (Sep,
  2014) 121101.

\bibitem{hoopercholis:2013}
I.~Cholis and D.~Hooper, {\it Dark matter and pulsar origins of the rising
  cosmic ray positron fraction in light of new data from the ams},  {\em Phys.
  Rev. D} {\bf 88} (Jul, 2013) 023013,
  [\href{http://arxiv.org/abs/astro-ph/1}{{\tt astro-ph/1}}].

\bibitem{profumolindenpulsars:2013}
T.~Linden and S.~Profumo, {\it Probing the pulsar origin of the anomalous
  positron fraction with ams-02 and atmospheric cherenkov telescopes},  {\em
  The Astrophysical Journal} {\bf 772} (2013), no.~1 18,
  [\href{http://arxiv.org/abs/astro-ph/1}{{\tt astro-ph/1}}].

\bibitem{yin:2013}
P.-F. Yin, Z.-H. Yu, Q.~Yuan, and X.-J. Bi, {\it Pulsar interpretation for the
  ams-02 result},  {\em Phys. Rev. D} {\bf 88} (Jul, 2013) 023001,
  [\href{http://arxiv.org/abs/astro-ph/1}{{\tt astro-ph/1}}].

\bibitem{weniger:2012}
C.~Weniger, {\it A tentative gamma-ray line from dark matter annihilation at
  the fermi large area telescope},  {\em Journal of Cosmology and Astroparticle
  Physics} {\bf 2012} (2012), no.~08 007,
  [\href{http://arxiv.org/abs/hep-ph/120}{{\tt hep-ph/120}}].

\bibitem{albert130gev:2014}
A.~Albert, {\it Indirect searches for dark matter with the fermi large area
  telescope},  {\em Physics Procedia} {\bf 61} (2015), no.~0 6--12. 13th
  International Conference on Topics in Astroparticle and Underground Physics,
  TAUP 2013.

\bibitem{Moduli:2013}
M.~Bose, M.~Dine, and P.~Draper, {\it Moduli or not},  {\em Physical Review D}
  {\bf 88} (2013) 023533, [\href{http://arxiv.org/abs/hep-ph/130}{{\tt
  hep-ph/130}}].

\bibitem{quintessence}
S.~Profumo and P.~Ullio, {\it Susy dark matter and quintessence},  {\em JCAP}
  {\bf 0311} (2003) 006, [\href{http://arxiv.org/abs/hep-ph/030}{{\tt
  hep-ph/030}}].

\bibitem{dmvelocity:2006}
C.~Savage, K.~Freese, and P.~Gondolo, {\it Annual modulation of dark matter in
  the presence of streams},  {\em Phys. Rev. D} {\bf 74} (Aug, 2006) 043531,
  [\href{http://arxiv.org/abs/astro-ph/0}{{\tt astro-ph/0}}].

\bibitem{dmvelocity:2004}
M.~Hoeft, J.~P. Mücket, and S.~Gottlöber, {\it Velocity dispersion profiles
  in dark matter halos},  {\em The Astrophysical Journal} {\bf 602} (2004),
  no.~1 162, [\href{http://arxiv.org/abs/astro-ph/0}{{\tt astro-ph/0}}].

\bibitem{feynarts:2001}
T.~Hahn, {\it Generating feynman diagrams and amplitudes with feynarts 3},
  {\em Computer Physics Communications} {\bf 140} (2001), no.~3 418 -- 431,
  [\href{http://arxiv.org/abs/hep-ph/001}{{\tt hep-ph/001}}].

\bibitem{formcalclooptools:1999}
T.~Hahn and M.~Pérez-Victoria, {\it Automated one-loop calculations in four
  and d dimensions},  {\em Computer Physics Communications} {\bf 118} (1999),
  no.~2–3 153 -- 165, [\href{http://arxiv.org/abs/hep-ph/980}{{\tt
  hep-ph/980}}].

\bibitem{oldloopa:1988}
L.~Bergstr\"om and H.~Snellman, {\it Observable monochromatic photons from
  cosmic photino annihilation},  {\em Physical Review D} {\bf 37} (Jun, 1988)
  3737--3741.

\bibitem{oldloopb:1989}
L.~Bergstr\"om, {\it Radiative processes in dark matter photino annihilation},
  {\em Physical Letters B} {\bf 225} (1989), no.~4 372.

\bibitem{oldloopc:1989}
S.~Rudaz, {\it Annihilation of heavy-neutral-fermion pairs into monochromatic
  $\gamma$ rays and its astrophysical implications},  {\em Physical Review D}
  {\bf 39} (Jun, 1989) 3549--3556.

\bibitem{neutralinogg:1995}
G.~Jungman and M.~Kamionkowski, {\it $\gamma$ rays from neutralino
  annihilation},  {\em Phys. Rev. D} {\bf 51} (Mar, 1995) 3121--3124,
  [\href{http://arxiv.org/abs/hep-ph/950}{{\tt hep-ph/950}}].

\bibitem{loopCorrection:1997}
L.~Bergstr\"om and P.~Ullio, {\it Full one-loop calculation of neutralino
  annihilation into two photons},  {\em Nuclear Physics B} {\bf 504} (1997),
  no.~1–2 27 -- 44, [\href{http://arxiv.org/abs/hep-ph/970}{{\tt
  hep-ph/970}}].

\bibitem{cahill-rowleyhedri:2015}
M.~Cahill-Rowley, S.~El~Hedri, W.~Shepherd, and D.~G.~E. Walker, {\it
  Perturbative unitarity constraints on charged/colored portals},
  \href{http://arxiv.org/abs/hep-ph/150}{{\tt hep-ph/150}}.

\bibitem{amsFits:2013a}
H.-B. Jin, Y.-L. Wu, and Y.-F. Zhou, {\it Implications of the first ams-02
  measurement for dark matter annihilation and decay},  {\em Journal of
  Cosmology and Astroparticle Physics} {\bf 2013} (2013), no.~11 026,
  [\href{http://arxiv.org/abs/hep-ph/130}{{\tt hep-ph/130}}].

\bibitem{amsFits:2013b}
J.~Kopp, {\it Constraints on dark matter annihilation from ams-02 results},
  {\em Phys. Rev. D} {\bf 88} (Oct, 2013) 076013,
  [\href{http://arxiv.org/abs/hep-ph/130}{{\tt hep-ph/130}}].

\bibitem{amsFits:2013c}
A.~D. Simone, A.~Riotto, and W.~Xue, {\it Interpretation of ams-02 results:
  correlations among dark matter signals},  {\em Journal of Cosmology and
  Astroparticle Physics} {\bf 2013} (2013), no.~05 003,
  [\href{http://arxiv.org/abs/hep-ph/130}{{\tt hep-ph/130}}].

\bibitem{amsfitsupdate:2014}
Q.-H. Cao, C.-R. Chen, and T.~Gong, {\it Leptophilic dark matter and ams-02
  cosmic-ray positron flux},  \href{http://arxiv.org/abs/hep-ph/140}{{\tt
  hep-ph/140}}.

\bibitem{hesslinesearch:2013}
A.~Abramowski, F.~Acero, F.~Aharonian, A.~G. Akhperjanian, G.~Anton,
  S.~Balenderan, A.~Balzer, A.~Barnacka, Y.~Becherini, J.~Becker~Tjus,
  K.~Bernl\"ohr, E.~Birsin, J.~Biteau, A.~Bochow, C.~Boisson, J.~Bolmont,
  P.~Bordas, J.~Brucker, F.~Brun, P.~Brun, T.~Bulik, S.~Carrigan, S.~Casanova,
  M.~Cerruti, P.~M. Chadwick, R.~C.~G. Chaves, A.~Cheesebrough,
  S.~Colafrancesco, G.~Cologna, J.~Conrad, C.~Couturier, M.~Dalton, M.~K.
  Daniel, I.~D. Davids, B.~Degrange, C.~Deil, P.~deWilt, H.~J. Dickinson,
  A.~Djannati-Ata\"\i, W.~Domainko, L.~O. Drury, G.~Dubus, K.~Dutson, J.~Dyks,
  M.~Dyrda, K.~Egberts, P.~Eger, P.~Espigat, L.~Fallon, C.~Farnier, S.~Fegan,
  F.~Feinstein, M.~V. Fernandes, D.~Fernandez, A.~Fiasson, G.~Fontaine,
  A.~F\"orster, M.~F\"u\ss{}ling, M.~Gajdus, Y.~A. Gallant, T.~Garrigoux,
  H.~Gast, B.~Giebels, J.~F. Glicenstein, B.~Gl\"uck, D.~G\"oring, M.-H.
  Grondin, S.~H\"affner, J.~D. Hague, J.~Hahn, D.~Hampf, J.~Harris, S.~Heinz,
  G.~Heinzelmann, G.~Henri, G.~Hermann, A.~Hillert, J.~A. Hinton, W.~Hofmann,
  P.~Hofverberg, M.~Holler, D.~Horns, A.~Jacholkowska, C.~Jahn, M.~Jamrozy,
  I.~Jung, M.~A. Kastendieck, K.~Katarzy\ifmmode~\acute{n}\else \'{n}\fi{}ski,
  U.~Katz, S.~Kaufmann, B.~Kh\'elifi, S.~Klepser, D.~Klochkov,
  W.~Klu\ifmmode~\acute{z}\else \'{z}\fi{}niak, T.~Kneiske, N.~Komin,
  K.~Kosack, R.~Kossakowski, F.~Krayzel, P.~P. Kr\"uger, H.~Laffon, G.~Lamanna,
  J.~Lefaucheur, M.~Lemoine-Goumard, J.-P. Lenain, D.~Lennarz, T.~Lohse,
  A.~Lopatin, C.-C. Lu, V.~Marandon, A.~Marcowith, J.~Masbou, G.~Maurin,
  N.~Maxted, M.~Mayer, T.~J.~L. McComb, M.~C. Medina, J.~M\'ehault, U.~Menzler,
  R.~Moderski, M.~Mohamed, E.~Moulin, C.~L. Naumann, M.~Naumann-Godo,
  M.~de~Naurois, D.~Nedbal, D.~Nekrassov, N.~Nguyen, J.~Niemiec, S.~J. Nolan,
  S.~Ohm, E.~de~O\~na Wilhelmi, B.~Opitz, M.~Ostrowski, I.~Oya, M.~Panter,
  R.~D. Parsons, M.~Paz~Arribas, N.~W. Pekeur, G.~Pelletier, J.~Perez, P.-O.
  Petrucci, B.~Peyaud, S.~Pita, G.~P\"uhlhofer, M.~Punch, A.~Quirrenbach,
  M.~Raue, A.~Reimer, O.~Reimer, M.~Renaud, R.~de~los Reyes, F.~Rieger,
  J.~Ripken, L.~Rob, S.~Rosier-Lees, G.~Rowell, B.~Rudak, C.~B. Rulten,
  V.~Sahakian, D.~A. Sanchez, A.~Santangelo, R.~Schlickeiser, A.~Schulz,
  U.~Schwanke, S.~Schwarzburg, S.~Schwemmer, F.~Sheidaei, J.~L. Skilton,
  H.~Sol, G.~Spengler, L.~Stawarz, R.~Steenkamp, C.~Stegmann, F.~Stinzing,
  K.~Stycz, I.~Sushch, A.~Szostek, J.-P. Tavernet, R.~Terrier, M.~Tluczykont,
  C.~Trichard, K.~Valerius, C.~van Eldik, G.~Vasileiadis, C.~Venter, A.~Viana,
  P.~Vincent, H.~J. V\"olk, F.~Volpe, S.~Vorobiov, M.~Vorster, S.~J. Wagner,
  M.~Ward, R.~White, A.~Wierzcholska, D.~Wouters, M.~Zacharias, A.~Zajczyk,
  A.~A. Zdziarski, A.~Zech, and H.-S. Zechlin, {\it Search for photon-linelike
  signatures from dark matter annihilations with h.e.s.s.},  {\em Phys. Rev.
  Lett.} {\bf 110} (Jan, 2013) 041301,
  [\href{http://arxiv.org/abs/astro-ph/1}{{\tt astro-ph/1}}].

\bibitem{fermilinesearch:2013}
M.~Ackermann, M.~Ajello, A.~Albert, A.~Allafort, L.~Baldini, G.~Barbiellini,
  D.~Bastieri, K.~Bechtol, R.~Bellazzini, E.~Bissaldi, E.~D. Bloom,
  E.~Bonamente, E.~Bottacini, T.~J. Brandt, J.~Bregeon, M.~Brigida, P.~Bruel,
  R.~Buehler, S.~Buson, G.~A. Caliandro, R.~A. Cameron, P.~A. Caraveo, J.~M.
  Casandjian, C.~Cecchi, E.~Charles, R.~C.~G. Chaves, A.~Chekhtman, J.~Chiang,
  S.~Ciprini, R.~Claus, J.~Cohen-Tanugi, J.~Conrad, S.~Cutini, F.~D'Ammando,
  A.~de~Angelis, F.~de~Palma, C.~D. Dermer, S.~W. Digel, L.~Di~Venere, P.~S.
  Drell, A.~Drlica-Wagner, R.~Essig, C.~Favuzzi, S.~J. Fegan, E.~C. Ferrara,
  W.~B. Focke, A.~Franckowiak, Y.~Fukazawa, S.~Funk, P.~Fusco, F.~Gargano,
  D.~Gasparrini, S.~Germani, N.~Giglietto, F.~Giordano, M.~Giroletti,
  T.~Glanzman, G.~Godfrey, G.~A. Gomez-Vargas, I.~A. Grenier, S.~Guiriec,
  M.~Gustafsson, D.~Hadasch, M.~Hayashida, A.~B. Hill, D.~Horan, X.~Hou, R.~E.
  Hughes, Y.~Inoue, E.~Izaguirre, T.~Jogler, T.~Kamae, J.~Kn\"odlseder,
  M.~Kuss, J.~Lande, S.~Larsson, L.~Latronico, F.~Longo, F.~Loparco, M.~N.
  Lovellette, P.~Lubrano, D.~Malyshev, M.~Mayer, M.~N. Mazziotta, J.~E.
  McEnery, P.~F. Michelson, W.~Mitthumsiri, T.~Mizuno, A.~A. Moiseev, M.~E.
  Monzani, A.~Morselli, I.~V. Moskalenko, S.~Murgia, T.~Nakamori, R.~Nemmen,
  E.~Nuss, T.~Ohsugi, A.~Okumura, N.~Omodei, M.~Orienti, E.~Orlando, J.~F.
  Ormes, D.~Paneque, J.~S. Perkins, M.~Pesce-Rollins, F.~Piron, G.~Pivato,
  S.~Rain\`o, R.~Rando, M.~Razzano, S.~Razzaque, A.~Reimer, O.~Reimer, R.~W.
  Romani, M.~S\'anchez-Conde, A.~Schulz, C.~Sgr\`o, J.~Siegal-Gaskins, E.~J.
  Siskind, A.~Snyder, G.~Spandre, P.~Spinelli, D.~J. Suson, H.~Tajima,
  H.~Takahashi, J.~G. Thayer, J.~B. Thayer, L.~Tibaldo, M.~Tinivella, G.~Tosti,
  E.~Troja, Y.~Uchiyama, T.~L. Usher, J.~Vandenbroucke, V.~Vasileiou,
  G.~Vianello, V.~Vitale, B.~L. Winer, K.~S. Wood, M.~Wood, Z.~Yang,
  G.~Zaharijas, and S.~Zimmer, {\it Search for gamma-ray spectral lines with
  the fermi large area telescope and dark matter implications},  {\em Phys.
  Rev. D} {\bf 88} (Oct, 2013) 082002,
  [\href{http://arxiv.org/abs/astro-ph/1}{{\tt astro-ph/1}}].

\bibitem{futurelinesearches:2012}
L.~Bergstr\"om, G.~Bertone, J.~Conrad, C.~Farnier, and C.~Weniger, {\it
  Investigating gamma-ray lines from dark matter with future observatories},
  {\em Journal of Cosmology and Astroparticle Physics} {\bf 2012} (2012),
  no.~11 025, [\href{http://arxiv.org/abs/hep-ph/120}{{\tt hep-ph/120}}].

\bibitem{cmb:2014}
M.~S. Madhavacheril, N.~Sehgal, and T.~R. Slatyer, {\it Current dark matter
  annihilation constraints from cmb and low-redshift data},  {\em Physical
  Review D} {\bf 89} (May, 2014) 103508,
  [\href{http://arxiv.org/abs/astro-ph.C}{{\tt astro-ph.C}}].

\bibitem{fermidiffuse:2012}
M.~Ackermann, M.~Ajello, W.~B. Atwood, L.~Baldini, G.~Barbiellini, D.~Bastieri,
  K.~Bechtol, R.~Bellazzini, R.~D. Blandford, E.~D. Bloom, E.~Bonamente, A.~W.
  Borgland, E.~Bottacini, T.~J. Brandt, J.~Bregeon, M.~Brigida, P.~Bruel,
  R.~Buehler, S.~Buson, G.~A. Caliandro, R.~A. Cameron, P.~A. Caraveo, J.~M.
  Casandjian, C.~Cecchi, E.~Charles, A.~Chekhtman, J.~Chiang, S.~Ciprini,
  R.~Claus, J.~Cohen-Tanugi, J.~Conrad, A.~Cuoco, S.~Cutini, F.~D’Ammando,
  A.~de~Angelis, F.~de~Palma, C.~D. Dermer, E.~do~Couto~e Silva, P.~S. Drell,
  A.~Drlica-Wagner, L.~Falletti, C.~Favuzzi, S.~J. Fegan, W.~B. Focke,
  Y.~Fukazawa, S.~Funk, P.~Fusco, F.~Gargano, D.~Gasparrini, S.~Germani,
  N.~Giglietto, F.~Giordano, M.~Giroletti, T.~Glanzman, G.~Godfrey, I.~A.
  Grenier, S.~Guiriec, M.~Gustafsson, D.~Hadasch, M.~Hayashida, D.~Horan, R.~E.
  Hughes, M.~S. Jackson, T.~Jogler, G.~Jóhannesson, A.~S. Johnson, T.~Kamae,
  J.~Knödlseder, M.~Kuss, J.~Lande, L.~Latronico, A.~M. Lionetto, M.~L. Garde,
  F.~Longo, F.~Loparco, B.~Lott, M.~N. Lovellette, P.~Lubrano, M.~N. Mazziotta,
  J.~E. McEnery, J.~Mehault, P.~F. Michelson, W.~Mitthumsiri, T.~Mizuno, A.~A.
  Moiseev, C.~Monte, M.~E. Monzani, A.~Morselli, I.~V. Moskalenko, S.~Murgia,
  M.~Naumann-Godo, J.~P. Norris, E.~Nuss, T.~Ohsugi, M.~Orienti, E.~Orlando,
  J.~F. Ormes, D.~Paneque, J.~H. Panetta, M.~Pesce-Rollins, M.~Pierbattista,
  F.~Piron, G.~Pivato, H.~Poon, S.~Rainò, R.~Rando, M.~Razzano, S.~Razzaque,
  A.~Reimer, O.~Reimer, C.~Romoli, C.~Sbarra, J.~D. Scargle, C.~Sgrò, E.~J.
  Siskind, G.~Spandre, P.~Spinelli, Łukasz Stawarz, A.~W. Strong, D.~J. Suson,
  H.~Tajima, H.~Takahashi, T.~Tanaka, J.~G. Thayer, J.~B. Thayer, L.~Tibaldo,
  M.~Tinivella, G.~Tosti, E.~Troja, T.~L. Usher, J.~Vandenbroucke,
  V.~Vasileiou, G.~Vianello, V.~Vitale, A.~P. Waite, E.~Wallace, K.~S. Wood,
  M.~Wood, Z.~Yang, G.~Zaharijas, and S.~Zimmer, {\it Constraints on the
  galactic halo dark matter from fermi-lat diffuse measurements},  {\em The
  Astrophysical Journal} {\bf 761} (2012), no.~2
  [\href{http://arxiv.org/abs/astro-ph.C}{{\tt astro-ph.C}}].

\bibitem{yang:1950}
C.~Yang, {\it Selection rules for the dematerialization of a particle into two
  photons},  {\em Physical Review} {\bf 77} (January, 1950) 242--245.

\bibitem{zgammagamma:2014}
E.~Zhemchugov, {\it On $z\to\gamma\gamma$ decay and cancellation of axial
  anomaly in $z\to\gamma\gamma$ transition amplitude for massive fermions},
  {\em Physics of Atomic Nuclei} {\bf 77} (2014), no.~11 1390--1399,
  [\href{http://arxiv.org/abs/hep-ph/140}{{\tt hep-ph/140}}].

\bibitem{colliderdips:2014}
Y.~Bai and W.-Y. Keung, {\it Dips at colliders},
  \href{http://arxiv.org/abs/hep-ph/140}{{\tt hep-ph/140}}.

\end{thebibliography}\endgroup

\newpage\clearpage
\appendix

\section{Analytic results}
\label{app:analytics}

\subsection{A selection of tree-level cross sections}
\label{app:tree-analytics}

Table \ref{tab:tree-analytics} shows the tree-level cross sections for several theories, evaluated to order $v_\chi^2$.

\begin{table}[h]
    \centering
    \begin{tabular}{|c|c|}
        \hline
		Theory & $\xsecff$ expanded to $\mathcal{O}(v_\chi^2)$\\
        \hline
        \rule[-17pt]{0pt}{40pt}SS & $\displaystyle \frac{\lambda^2 v_\chi^2}{4\pi}(M_\chi^2 - m_f^2) \left( 1 - \frac{m_f^2}{M_\chi^2} \right)^{1/2}$\\
        \hline
        \rule[-17pt]{0pt}{40pt}PP & $\displaystyle \frac{\lambda^2 M_\chi^2}{4\pi} \left( 1 - \frac{m_f^2}{M_\chi^2} \right)^{1/2} + \frac{\lambda^2v^2}{8\pi} (2M_\chi^2 - m_f^2) \left( 1 - \frac{m_f^2}{M_\chi^2} \right)^{-1/2}$\\
        %\hline
        %\rule[-17pt]{0pt}{40pt}VV & $\displaystyle \frac{\lambda^2}{4\pi}(2M_\chi^2 + m_f^2) \left( 1 - \frac{m_f^2}{M_\chi^2} \right)^{1/2} + \frac{\lambda^2 v_\chi^2}{48\pi M_\chi^2} (13M_\chi^4 - 2M_\chi^2 m_f^2 + 7m_f^4) \left( 1 - \frac{m_f^2}{M_\chi^2} \right)^{-1/2}$\\
        \hline
        %\rule[-17pt]{0pt}{40pt}VA & $\displaystyle \frac{\lambda^2}{2\pi} (M_\chi^2 - m_f^2) \left( 1 - \frac{m_f^2}{M_\chi^2} \right)^{1/2} + \frac{\lambda^2 v_\chi^2}{48\pi} (13M_\chi^2 + 23 m_f^2) \left( 1 - \frac{m_f^2}{M_\chi^2} \right)^{1/2}$\\
        %\hline
        \rule[-17pt]{0pt}{40pt}AV & $\displaystyle \frac{\lambda^2 v_\chi^2}{48\pi} (13M_\chi^2 + 11m_f^2) \left( 1 - \frac{m_f^2}{M_\chi^2} \right)^{1/2}$\\
        \hline
        \rule[-17pt]{0pt}{40pt}AA & $\displaystyle \frac{\lambda^2 m_f^2}{4\pi} \left( 1 - \frac{m_f^2}{M_\chi^2} \right)^{1/2} + \frac{\lambda^2 v_\chi^2}{48\pi M_\chi^2} (13M_\chi^4 - 38M_\chi^2 m_f^2 + 31m_f^2) \left( 1 - \frac{m_f^2}{M_\chi^2} \right)^{-1/2}$\\
        \hline
        %\rule[-17pt]{0pt}{40pt}tchan.mod & $\displaystyle \frac{\sqrt{M_\chi^2 - m_f^2}[(\lambda_L^2 + \lambda_R^2) m_f + 2\lambda_L \lambda_R M_\chi] [({\lambda_L^*}^2 + {\lambda_R^*}^2)m_f + 2 \lambda_L^* \lambda_R^* M_\chi]}{64\pi M_\chi M_{\tilde{f}}^4} + \bigo(v_\chi^2)$\\
        %\hline
    \end{tabular}
    \caption{Tree-level cross sections for a representative subset of EFTs.}
    \label{tab:tree-analytics}
\end{table}

\begin{figure}
    \centering
        \begin{fmffile}{triangles}
            \begin{fmfgraph*}(130,50)
                \fmfstraight
                \fmftop{pt1,pt2,pt3,pt4}
                \fmfbottom{pb1,pb2,pb3,pb4}
                \fmfleft{chi,chibar}
                \fmfright{gamma2,gamma1}
                \fmf{fermion,label=$k_1$}{chi,v}
                \fmf{fermion,label=$k_2$}{v,chibar}
                \fmflabel{$\chi$}{chi}
                \fmflabel{$\bar{\chi}$}{chibar}
                \fmfblob{25}{v}
                \fmf{fermion}{g2rad,v,g1rad}
                \fmf{fermion,label=$p$}{g1rad,g2rad}
                \fmf{phantom,tension=5}{pt3,g1rad}
                \fmf{phantom,tension=5}{pb3,g2rad}
                \fmfv{label=$\mu$,label.dist=10}{g1rad}
                \fmfv{label=$\nu$,label.dist=10}{g2rad}
                \fmf{photon,tension=1,label=$k_3$}{g1rad,gamma1}
                \fmf{photon,tension=1,label=$k_4$}{g2rad,gamma2}
                \fmflabel{$\gamma$}{gamma1}
                \fmflabel{$\gamma$}{gamma2}
            \end{fmfgraph*}
        \end{fmffile}
        \hspace{0.5in}
        \begin{fmffile}{trianglescrossed}
            \begin{fmfgraph*}(130,50)
                \fmfstraight
                \fmftop{pt1,pt2,pt3,pt4}
                \fmfbottom{pb1,pb2,pb3,pb4}
                \fmfleft{chi,chibar}
                \fmflabel{$\chi$}{chi}
                \fmflabel{$\bar{\chi}$}{chibar}
                \fmf{fermion,label=$k_1$}{chi,v}
                \fmf{fermion,label=$k_2$}{v,chibar}
                \fmfright{gamma2,gamma1}
                \fmfblob{25}{v}
                \fmf{fermion}{g2rad,v,g1rad}
                \fmf{fermion,label=$p$}{g1rad,g2rad}
                \fmf{phantom,tension=5}{pt3,g1rad}
                \fmf{phantom,tension=5}{pb3,g2rad}
                \fmflabel{$\mu$}{g1rad}
                \fmflabel{$\nu$}{g2rad}
                \fmf{photon,tension=0,label=$k_3$,label.dist=20}{g2rad,gamma1}
                \fmf{phantom,tension=1}{g1rad,gamma1}
                \fmf{photon,tension=0,label=$k_4$,label.dist=20}{g1rad,gamma2}
                \fmf{phantom,tension=1}{g2rad,gamma2}
                \fmflabel{$\gamma$}{gamma1}
                \fmflabel{$\gamma$}{gamma2}
            \end{fmfgraph*}
        \end{fmffile}
    \caption{Triangle diagrams contributing to the process $\chi\bar{\chi}\to\gamma\gamma$.}
    \label{fig:triangles}
\end{figure}
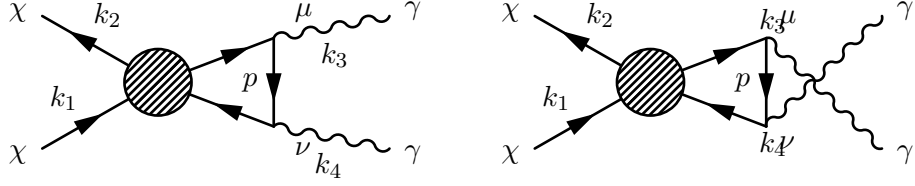

\subsection{Loop cross section for SP theory}
\label{app:loop-analytics-SP}

The total loop amplitude is found by summing the diagrams shown in figure \ref{fig:triangles}:
\begin{align}
    \M_\u{loop}^\u{SP} &= \lambda [\bar{v_\chi}^r(k_2) u^s(k_1)] \varepsilon^*_\mu(k_3) \varepsilon^*_\nu(k_4) N^{\mu\nu}(k_3,k_4),
\end{align}
where $\lambda$ is the coupling at the blob and $N^{\mu\nu}(k_3,k_4) = \tilde{N}^{\mu\nu}(k_3,k_4) + \tilde{N}^{\nu\mu}(k_4,k_3)$ is the sum of loop integrals for the crossed and uncrossed diagrams.  The former is given by
\begin{align}
    \tilde{N}^{\mu\nu}(k_3,k_4) &= (-ie)^2\int \frac{d^4p}{(2\pi)^4} \frac{\tr[(\s p - \s k_4 + m_f)\gamma^\nu(\s p + m_f)\gamma^\mu(\s p + \s k_3 + m_f)\gamma^5]}{[(p - k_4)^2 - m_f^2][p^2 - m_f^2][(p + k_3)^2 - m_f^2]}.
\end{align}
The trace evaluates to $4 i m_f \varepsilon^{\mu\nu\alpha\beta} {k_3}_\alpha {k_4}_\beta$ by standard identities.  After introducing Feynman parameters, the integral becomes
\begin{align}
    \tilde{N}^{\mu\nu}(k_3,k_4) &= \frac{i e^2 m_f}{4\pi^4} \varepsilon^{\mu\nu\alpha\beta} {k_3}_\alpha {k_4}_\beta \int_0^1 dx\ \int_0^{1-x} dz\ \int \frac{d^4l}{[l^2 - m_f^2 + 2 x z k_3\cdot k_4]^3},\ l \equiv p - xk_4 + zk_3.
\end{align}
Wick rotating and integrating yields
\begin{align}
    \tilde{N}^{\mu\nu}(k_3,k_4) &= -\frac{i e^2 m_f}{4\pi^2 s} \varepsilon^{\mu\nu\alpha\beta} {k_3}_\alpha {k_4}_\beta \left[ \li_2\left( \frac{2s}{s - \sqrt{s(s-4m_f^2)}} \right) + \li_2\left( \frac{2s}{s+\sqrt{s(s-4m_f^2)}} \right) \right].
\end{align}
This can be converted to the more pleasant expression found in [CITE] by applying Landen's identity:
\begin{align}
	\tilde{N}^{\mu\nu}(k_3,k_4) &= \frac{i e^2 m_f}{8\pi^2 s} \varepsilon^{\mu\nu\alpha\beta} {k_3}_\alpha {k_4}_\beta  \log^2 \left( 1 - \frac{s + \sqrt{s(s-4m_f^2)}}{2m_f^2} \right).
\end{align}
Since this expression is symmetric under $(\mu\leftrightarrow\nu,k_3\leftrightarrow k_4)$, the total amplitude is found to be
\begin{align}
	\M_\u{loop}^\u{SP} &=  \frac{i \lambda e^2 m_f}{4\pi^2 s} [\bar{v_\chi}^r(k_2) u^s(k_1)] \varepsilon^{\mu\nu\alpha\beta} \varepsilon^*_\mu(k_3) \varepsilon^*_\nu(k_4) {k_3}_\alpha {k_4}_\beta  \log^2\left( 1 - \frac{s + \sqrt{s(s-4m_f^2)}}{2m_f^2} \right).
\end{align}
The spin and polarization-averaged matrix element is therefore
\begin{align}
    |\M_\u{loop}^\u{SP}|^2 &= \frac{\lambda^2 \alpha^2 m_f^2}{4\pi^2} (s - 4m_f^2) \log^2\left( 1 - \frac{s + \sqrt{s(s-4m_f^2)}}{2m_f^2} \right).
\end{align}
Doing the Lorentz invariant phase space integration gives the final result for the cross section:
\begin{align}
    \xsecgg^\u{SP} &= \frac{\lambda^2 \alpha^2 m_f^2}{256 \pi^4} \left( 1 - \frac{4m_f^2}{s} \right) \left| \log^2\left( 1 - \frac{s + \sqrt{s(s-4m_f^2)}}{2m_f^2} \right) \right|^2.
\end{align}

\subsection{Loop cross section for AA theory}
\label{app:loop-analytics-AA}

Using the result for $\M_{Z\to\gamma\gamma}$ from [CITE], we find
\begin{align}
    \xsecgg &= \frac{\lambda^2 \alpha^2 M_\chi^2 s^2}{4\pi^3} \left| 1 + \frac{m_f^2}{s} \log^2\left( 1 - \frac{s + \sqrt{s(s - 4m_f^2)}}{2m_f^2} \right) \right|^2.
\end{align}
Note that equations 3.24 and 3.25 in [CITE] are both missing a factor of $M_Z^{-1}$.

\section{Yang's theorem}
\label{app:yang}
Yang's theorem \cite{yang:1950} states that a massive spin-1 particle cannot decay into a pair of identical massless spin-1 particles when the particles are all on shell, which has implications about $\xsecgg$ for vector-type EFTs.  Any of our effective operators can be interpreted as the limit of a $Z'$ theory as $M_{Z'} \to \infty$.  Here we will use this perspective to make some remarks about the process $\chi\chi\to\gamma\gamma$ for vector-like theories of the form $\biL{\Gamma^\mu}{\Sigma_\mu}$.

For such a theory, the EFT amplitude for this process can be written schematically as
\begin{align}
	\M^{\u{EFT}}_{\chi\chi\to\gamma\gamma}(q) &= [\bar{v} \Gamma_\mu u] \cdot g^{\mu\nu} \cdot N_{\nu\rho\sigma} \cdot \varepsilon_\rho^* \varepsilon_\sigma^*,
\end{align}
where $N_{\nu\rho\sigma}$ represents the loop piece of $\M^{\u{EFT}}_{\chi\chi\to\gamma\gamma}$.  This is just the limit of the following amplitude in a $Z'$ theory:
\begin{align}
	\M_{\chi\chi\to\gamma\gamma}(q) &= [\bar{v} \Gamma_\mu u] \cdot \frac{1}{q^2 - M_{Z'}^2} \left( g^{\mu\nu} - \frac{q^\mu q^\nu}{M_{Z'}^2} \right) \cdot N_{\nu\rho\sigma} \cdot \varepsilon_\rho^* \varepsilon_\sigma^*,
\end{align}
where $q$ is the sum of the WIMP momenta.  When the $Z'$ is on shell, the factor in the propagator with Lorentz indices is the projection onto the its physical polarizations, $-\sum_{\varepsilon^\mu q_\mu = 0} \varepsilon^\mu(q) {\varepsilon^\nu}^*(q)$.  Thus Yang's theorem requires that $(q^2 - M_{Z'}^2) \M_{\chi\chi\to\gamma\gamma}$ vanish when $q = \sqrt{s} = M_{Z'}$.

There are two ways this can happen.  When $\M_{\chi\chi\to\gamma\gamma}(M_{Z'}) = 0$, this condition will of course be satisfied.  This is true when $\Sigma^\mu = \gamma^\mu$, where Furry's theorem gives the much stronger result that $N_{\nu\rho\sigma}$ and thus $\M_{\chi\chi\to\gamma\gamma}(q) = 0$ for all $q$.  $|\M_{\chi\chi\to\gamma\gamma}(q)|^2$ also vanishes for the VA theory.  % I don't understand why this vanishes!!!  See calculation in va_gammagamma_check.nb.
The only other way this can happen is if $\M_{\chi\chi\to\gamma\gamma}(q)$ contains a factor of $q^2 - M_{Z'}^2$ to cancel the one in the propagator.  This only occurs when the $Z'$ has nonzero axial coupling to the WIMP and fermion.  We have verified this using FeynCalc and the analytic expression for $Z'\to\gamma\gamma$ \cite{zgammagamma:2014}.

However, \cite{zgammagamma:2014} demonstrated that the triangle diagram for $Z'\to\gamma\gamma$ gives rise to an anomaly, which is cancelled by mixing between the Goldstone bosons arising from spontaneous symmetry breaking and the $Z'$.  \emph{This means that $\xsecgg$ vanishes for $Z'$ models.}\footnote{Since the matrix element in this cross section is zero, the third case discussed in \cite{colliderdips:2014} cannot occur.}  Of course, effective operators with axial coupling to the WIMP and fermion can arise in theories such as the $t$-channel one we considered in section \ref{subsec:tchan}.  In this case there is no anomaly since no gauge symmetry is being violated, and $\xsecgg$ is nonvanishing.

\end{document}